\documentclass[12pt]{article}
\pdfoutput=1

\usepackage{putex}
\usepackage{graphicx}
\usepackage{caption}
\usepackage{amsmath}
\usepackage{array}
\usepackage{subcaption}
\usepackage{epstopdf}
\usepackage{enumerate}
\usepackage{cite}
\usepackage{youngtab}
\usepackage{tensor}
\usepackage{slashed}
\usepackage[aligntableaux=center]{ytableau}
\usepackage[utf8]{inputenc}
\usepackage{rotating}
\usepackage{bigfoot}
\usepackage[
      colorlinks=true,
      linkcolor=blue,
      urlcolor=blue,
      filecolor=black,
      citecolor=red,
      ]{hyperref}

\newcommand{\abs}[1]{\left\lvert #1 \right\rvert}

\newcommand {\be} {\begin {equation}}
\newcommand {\ee} {\end {equation}}

\newcommand {\bes} {\begin {equation*}}
\newcommand {\ees} {\end {equation*}}

\newcommand{\es}[2] {\begin{equation} \label{#1} \begin{split} #2 \end{split} \end{equation}}

\newcommand{\cA}{{\mathcal A}}

\newcommand{\cF}{{\mathcal F}}
\newcommand{\cG}{{\mathcal G}}

\newcommand{\cL}{{\mathcal L}}

\newcommand{\cO}{{\mathcal O}}

\newcommand{\cS}{{\mathcal S}}
\newcommand{\cT}{{\mathcal T}}

\newcommand{\cM}{{\mathcal M}}

\newcommand{\beq}{\begin{equation}}
\newcommand{\eeq}{\end{equation}}

\def\ie{\begin{equation}\begin{aligned}}
\def\fe{\end{aligned}\end{equation}}

\newcommand{\n}{\nu}

\numberwithin{equation}{section}

\def\<{\langle}
\def\>{\rangle}

\begin{document}

\institution{Exile}{Department of Particle Physics and Astrophysics, Weizmann Institute of Science, Rehovot, Israel}

\title{Genus-2 Holographic Correlator on $AdS_5 \times S^5$ from Localization}

\authors{Shai M.~Chester\worksat{\Exile}}

\abstract{
We consider the four-point function of the stress tensor multiplet superprimary in $\mathcal{N}=4$ super-Yang-Mills (SYM) with gauge group $SU(N)$ in the large $N$ and large 't Hooft coupling $\lambda\equiv g_\text{YM}^2N$ limit, which is holographically dual to the genus expansion of IIB string theory on $AdS_5\times S^5$. In \cite{Binder:2019jwn} it was shown that the integral of this correlator is related to derivatives of the mass deformed $\mathcal{N}=2^*$ sphere free energy, which was computed using supersymmetric localization to leading order in $1/N^2$ for finite $\lambda$. We generalize this computation to any order in $1/N^2$ for finite $\lambda$ using topological recursion, and use this any order constraint to fix the $R^4$ correction to the holographic correlator to any order in the genus expansion. We also use it to complete the derivation of the 1-loop supergravity correction, and show that analyticity in spin fails at zero spin in the large $N$ expansion as predicted from the Lorentzian inversion formula. In the flat space limit, the $R^4$ term in the holographic correlator matches that of the IIB S-matrix in 10d, which is a precise check of AdS$_5$/CFT$_4$ for local operators at genus-one. Using the flat space limit and localization we then fix $D^4R^4$ in the holographic correlator to any order in the genus expansion, which is nontrivial at genus-two, i.e. $1/N^6$. This is the first result at two orders beyond the planar limit at strong coupling for a holographic correlator.
}
\date{\today}

\maketitle

\tableofcontents

\section{Introduction}
\label{intro}

The graviton S-matrix is one of the simplest observables in flat space quantum gravity. In IIB string theory in ten flat dimensions, the S-matrix can be computed for small string coupling $g_s$ and finite string length $\ell_s$ using a genus expansion of the worldsheet. When IIB string theory is compactified on $AdS_5\times S^5$, the scattering of gravitons is holographically dual to four point functions of single-trace half-BPS operators in the boundary CFT, which is maximally supersymmetric $\mathcal{N}=4$ super-Yang-Mills (SYM) with gauge group $SU(N)$ \cite{Maldacena:1997re}. In the 't Hooft limit where $N\to\infty$ for fixed 't Hooft coupling $\lambda=g_\text{YM}^2N$ and then $\lambda\to\infty$, the CFT correlators can be computed using classical IIB supergravity on $AdS_5\times S^5$, where $1/N^2$ corrections correspond to higher genus corrections in string theory and $1/\lambda$ corrections correspond to higher derivative corrections to supergravity. Unlike the flat space S-matrix, there is no systematic way to compute all these terms. At leading order in $1/\lambda$ and $1/N$, i.e. tree level supergravity, the correlators can be computed using Witten diagrams \cite{Eden:2000bk,Arutyunov:2002fh,Arutyunov:2003ae,Berdichevsky:2007xd,Uruchurtu:2008kp,Uruchurtu:2011wh,Arutyunov:2017dti,Arutyunov:2018tvn}, but this becomes difficult at loop level or for higher derivative corrections to supergravity since the contact terms are not fully known (see however \cite{deHaro:2002vk,Policastro:2006vt,Paulos:2008tn,Liu:2013dna} for partial results). Recently, some of these terms have been computed using supersymmetric localization \cite{Pestun:2007rz,Binder:2019jwn}, the flat space limit \cite{Polchinski:1999ry,Susskind:1998vk,Giddings:1999jq,Fitzpatrick:2011hu,Penedones:2010ue,Fitzpatrick:2011jn,Goncalves:2014ffa}, and unitarity methods \cite{Aharony:2016dwx,Alday:2017xua,Aprile:2017bgs,Aprile:2017xsp,Aprile:2017qoy,Alday:2018pdi,Alday:2018kkw,Drummond:2019odu,Alday:2017vkk}, which do not require an explicit bulk action as in the original  analytic bootstrap\footnote{See \cite{Caron-Huot:2018kta,Goncalves:2019znr,Rastelli:2017ymc,Rastelli:2019gtj,Zhou:2017zaw,Zhou:2018ofp,Alday:2014tsa,Chester:2018lbz,Chester:2018aca,Chester:2018dga,Binder:2018yvd,Binder:2019mpb,Aprile:2018efk,Giusto:2018ovt,Giusto:2019pxc,Alday:2017gde,Alday:2019qrf} for other applications of these methods to holographic correlators in various dimensions.} paper \cite{Heemskerk:2009pn} as first applied to $\mathcal{N}=4$ SYM in  \cite{Rastelli:2017udc}. In this work, we will use an extension of the localization method of \cite{Binder:2019jwn} along with the flat space limit to fix more of these terms, including the genus-one $R^4$ and the genus-two $D^4R^4$ contact terms, as well as the complete 1-loop supergravity term.

Before we discuss the expansion of the holographic correlator in $AdS_5\times  S^5$, it is simpler to consider the IIB S-matrix that is related to this correlator in the flat space limit. The IIB S-matrix has been computed in a small $g^2_s$ expansion to genus-two for finite $\ell_s$ \cite{Gomez:2010ad,DHoker:2005jhf}, and to genus-three \cite{Gomez:2013sla} to the lowest few orders in $\ell_s$. We will consider the following terms in the small $g_s$ and $\ell_s$ expansion of this amplitude:
\es{A}{
\cA=\cA_\text{SG}& \left[\left(1+\ell_s^6f^0_{R^4}(s,t)+\ell_s^{10}f^0_{D^4R^4}(s,t)+O(\ell_s^{12}) \right)\right.\\
&\left.+g_s^2\left( \ell_s^6f^1_{R^4}(s,t)+\ell_s^8f^1_{\text{SG}|\text{SG}}(s,t)+\ell_s^{10}f^1_{D^4R^4}(s,t)+O(\ell_s^{12}) \right)\right.\\
&\left.+g_s^4\left(\ell_s^6f^2_{R^4}(s,t)+\ell_s^{10}f^2_{D^4R^4}+O(\ell_s^{12}) \right)+O(g_s^6)\right]\,,
}
where we normalized the amplitude by the genus-zero supergravity term $\mathcal{A}_\text{SG}$, and $s,t,u=-s-t$ are Mandelstam variables. Higher orders in $\ell_s$ can come from contact terms of higher derivative correction to supergravity, which are analytic in $s,t,u$ and have an expansion in $g_s$, as well as loops, which are non-analytic in $s,t,u$. The first couple higher derivative terms are $R^4$ and $D^4R^4$. These terms are protected, and so only receive corrections at genus-zero as well as genus-one and two for $R^4$ and $D^4R^4$, respectively, which take the form \cite{Polchinski:1998rr,GREEN1982444}
\es{fs}{
&f_{R^4}^0=\frac{\zeta(3)}{32}stu\,,\qquad f^0_{D^4R^4}=\frac{\zeta(5)}{2^{10}}stu(s^2+t^2+u^2)\,,\\
& f^1_{R^4}=\frac{\pi^2}{96}stu\,,\qquad\;\; \;f^2_{D^4R^4}=\frac{\pi^4}{2^{9}\cdot135}stu(s^2+t^2+u^2)\,.
}
The only loop term shown in \eqref{A} is the 1-loop term with two supergravity vertices, which can be computed in terms of genus-zero supergravity using unitarity \cite{Green:2008uj}.

On $AdS_5\times S^5$ with AdS radius $L$, we consider the scattering of scalars in the supergraviton multiplet, which is holographically dual to the $\mathcal{N}=4$ SYM correlator $\langle SSSS\rangle$, where $S$ is the bottom component of the stress tensor multiplet, and is a scalar with dimension 2 that transforms in the $\bf{20}'$ of the $SU(4)$ R-symmetry. Superconformal Ward identities fix this correlator in terms of a single function of the conformal cross ratios \cite{Dolan:2001tt}, whose Mellin transform \cite{Penedones:2010ue,Fitzpatrick:2011ia} we denote by $\cM(s,t)$. In the strong coupling 't Hooft limit, $\cM$ has an expansion in $1/N$ and $1/\lambda$, whose scaling we can determine from \eqref{A} using the AdS/CFT dictionary
 \es{dict}{
 \frac{L^4}{\ell^4_{s}}= \lambda = {g_\text{YM}^2 N}\,, \qquad
  g_s = \frac{g_\text{YM}^2}{4\pi} \,.
 }
 Crossing symmetry and the analytic structure of Witten diagrams in Mellin space \cite{Penedones:2010ue,Fitzpatrick:2011ia,Fitzpatrick:2011hu,Fitzpatrick:2011dm} fix the allowed terms in this expansion to be\footnote{Our notation for polynomial Mellin amplitudes is related to \cite{Binder:2019jwn} as $\cM^{n}_\text{here}=\cM^{n+4}_\text{there}$. The difference in four is more natural for the reduced correlator, where polynomial Mellin amplitudes are 4 degrees less than the same term in the full correlator.}
\es{MellinIntro}{
   \cM =&\frac1 c\left[8\cM^\text{SG}+\lambda^{-\frac32}B_0^0 \cM^{0}  +\lambda^{-\frac52}\left[B_{2}^2 \cM^{2}+B_0^2\cM^{0}\right]+O(\lambda^{-3})\right]\\
   &+\frac{1}{c^2}
   \left[\lambda^{\frac12} \overline{B}_0^0 \cM^{0}+\left[\mathcal{M}^{\text{SG}|\text{SG}}+\overline{B}^{\text{SG}|\text{SG}}_0\mathcal{M}^0\right] +\lambda^{-\frac12}\left[ \overline{B}_2^2 \cM^{2} + \overline{B}_0^2 \cM^{0}  \right]+O(\lambda^{-1})\right]\\
   &+\frac{1}{c^3}
   \left[\lambda^{\frac52} \overline{\overline{B}}_0^0 \cM^{0}+\lambda^{\frac32}\left[ \overline{\overline{B}}_2^2 \cM^{2} + \overline{\overline{B}}_0^2 \cM^{0}  \right]+O(\lambda)\right]+O(c^{-4})\,.
}
where $c = (N^2 - 1)/4$ is the $c$ anomaly coefficient, which is the natural expansion for holographic correlators since it is simply related to the 5d Newton's constant. The Mellin amplitudes are functions of the Mellin variables $s$ and $t$, which are related to the Mandelstam variables in \eqref{A} in the flat space limit. As in the flat space S-matrix, we have degree $p$ in $s,t,u$ Mellin amplitudes $\cM^p$ that correspond to contact Witten diagrams with higher derivative corrections, as well as the 1-loop supergravity Mellin amplitude $\mathcal{M}^{\text{SG}|\text{SG}}$ that is non-analytic in $s,t,u$. The coefficient of the supergravity term is fixed by a Ward identity from the conservation of the stress tensor, and the non-analytic 1-loop term is fixed in terms of tree level supergravity using unitarity \cite{Aprile:2017bgs,Alday:2017xua,Alday:2018kkw} up to a constant ambiguity $\cM^0$. 

The remaining coefficients in \eqref{MellinIntro} have been addressed using two methods. Firstly, in the flat space limit \eqref{MellinIntro} can be related to the IIB S-matrix \eqref{A}, which was originally used in \cite{Goncalves:2014ffa} to fix the genus-zero $R^4$ term. In \cite{Binder:2019jwn}, this term was also fixed by relating the integral of $\langle SSSS\rangle$ to derivatives $\partial_m^2\partial_{\lambda^{-1}}F\vert_{m=0}$ of the mass $m$ deformed $\mathcal{N}=2^*$ free energy $F$ on $S^4$, which can be computed from a matrix model using supersymmetric localization \cite{Pestun:2007rz}. This quantity was computed to leading order in the 't Hooft limit in \cite{Russo:2013kea}, and used along with the flat space limit in \cite{Binder:2019jwn} to fix the coefficients of both $R^4$ and $D^4R^4$ at genus-zero to get
\es{oldBs}{
B_0^0=120\zeta(3)\,,\qquad B_2^2=630\zeta(5)\,,\qquad B_0^2=-1890\zeta(5)\,.
}

To fix coefficients beyond genus-zero, $\partial_m^2\partial_{\lambda^{-1}}F\vert_{m=0}$ must be computed to higher orders. In 3d, the holographic correlator in ABJM theory \cite{Aharony:2008ug} with gauge group $U(N)_k\times U(N)_{-k}$ and Chern-Simons level $k$, which is dual to IIA string theory on $AdS_4\times \mathbb{CP}_3$ in the large $N$ and large $\lambda=N/k$ expansion, was also related to derivatives of the mass deformed $S^3$ free energy \cite{Binder:2018yvd}. This quantity was computed using localization \cite{Kapustin:2009kz} in terms of a matrix model that is quite complicated, but can still be expanded to all order in $1/N$ using the Fermi gas method \cite{Marino:2011eh,Nosaka:2015iiw}. This all orders constraint could then be used to fix the $R^4$ correction to all orders in the genus expansion \cite{Binder:2019mpb}.\footnote{The $R^4$ correction has also been computed in the finite $k$ ABJM theory \cite{Chester:2018aca} and the 6d $(2,0)$ theory \cite{Chester:2018dga}, which are both dual to M-theory.} In our 4d case, no such Fermi gas method exists for the matrix model of the mass deformed $\mathcal{N}=2^*$ sphere free energy $F$, which also takes a complicated form. Instead, in this work we take advantage of the fact that the $m=0$ free energy for $\mathcal{N}=4$ SYM is just a free Gaussian matrix model, so $\partial_m^2\partial_{\lambda^{-1}}F\vert_{m=0}$ can be written as an expectation value in this free theory. This expectation value can then be computed to any order in $1/N$ at finite or perturbative $\lambda$ using topological recursion \cite{Eynard:2004mh,Eynard:2008we}, and also at finite $N$ and $\lambda$ (if we ignore non-perturbative instantons in the Nekrasov partition function) using orthogonal polynomials \cite{mehta1981}. 

We then use this any order in $1/N$ method to fix the coefficient of $\cM_0$ to any genus. For $R^4$, we compute the coefficient $\overline{B}_0^0$ of the genus-one correction, which scales as $\sqrt{\lambda}c^{-2}$, and is the only nonzero correction to $R^4$ beyond genus-zero. We can also fix the coefficient $\overline{B}_0^{\text{SG}|\text{SG}}$ of the constant ambiguity in the 1-loop supergravity term, which completes the derivation of \cite{Aprile:2017bgs,Alday:2017xua,Alday:2018kkw}. This constant term only contributes to scalar CFT data. In \cite{Aprile:2017bgs}, it was conjectured that $\overline{B}_0^{\text{SG}|\text{SG}}=0$\footnote{This coefficient was called $\frac{\alpha}{16}$ in that paper, since they expand in $4c=N^2-1$.} so that the anomalous dimension of the lowest twist double-trace operator would be analytic in spin down to zero spin. We find that $\overline{B}_0^{\text{SG}|\text{SG}}\neq0$, so analyticity in spin fails at zero spin, as expected from the Lorentzian inversion formula in the large $N$ limit of SYM \cite{Caron-Huot:2017vep}.

We also use the flat space limit and the known IIB S-matrix to fix the leading $s,t$ contribution to each Mellin amplitude. For $R^4$, we get the same answer for the genus-one term, which is the first genus-one check of AdS$_5/$ CFT$_4$ for local operators that could not be determined from genus-zero.\footnote{In \cite{Alday:2017vkk,Alday:2018pdi,Alday:2018kkw} various non-analytic genus-one terms in the holographic correlator were matched to the corresponding non-analytic term in the IIB S-matrix, but in both flat space and $AdS_5\times S^5$ these quantities are completely fixed by genus-zero data.} We are also able to fix $D^4R^4$ at genus-two, which is the only nonzero correction to $D^4R^4$ beyond genus-zero, by combining the two constraints from the flat space limit and localization. This is the first correction to a holographic correlator computed at genus-two, which is two orders beyond the planar limit.

The rest of this paper is organized as follows.  In Section~\ref{4point}, we review properties of the stress tensor multiplet four-point function in the strong coupling limit, including the flat space limit and the relation to the $\mathcal{N}=2^*$ sphere free energy. In Section~\ref{freem}, we show how topological recursion can be used to efficiently compute this quantity to any order in $1/N^2$, which we explicitly carry out to the lowest five orders, and check for finite $N$ using the method of orthogonal polynomials. In Section \ref{relate}, we use this all orders constraint to fix the higher genus coefficients in the holographic correlator. We conclude with a discussion in Section~\ref{conc}. We include many explicit results from the localization calculation in the appendices.

\section{$\mathcal{N}=4$ stress-tensor four-point function}
\label{4point}

We begin by reviewing what is already known about the stress tensor multiplet four-point function. First we discuss general constraints from the $\mathcal{N}=4$ superconformal group. Then we discuss Mellin space, the large $N$ strong coupling expansion for SYM, and the flat space limit. Lastly, we review the relation derived in \cite{Binder:2019jwn} between the integrated stress tensor correlator and the $\mathcal{N}=2^*$ free energy on $S^4$.

\subsection{Setup}
\label{setup}

Let us denote the bottom component of the stress tensor multiplet as $S$, which is a dimension 2 scalar in the $\bf20'$ of the $SU(4)$ R-symmetry. We can express this operator as a traceless symmetric tensor $S_{IJ}$ of $SO(6)\cong SU(4)$ fundamental indices $I,J=1,\dots,6$. To avoid carrying around indices, it is convenient to contract them with auxiliary polarization vectors $Y^I$ that are constrained to be null, i.e. $Y\cdot Y=0$, so that we define
\es{polarization}{
S(\vec x,Y)\equiv S_{IJ}Y^IY^J\,,
}
where $\vec x$ denotes the position dependence. We normalize $S$ so that its two-point function is
\es{2point}{
\langle S(\vec x_1,Y_1)S(\vec x_2,Y_2)\rangle=\frac{Y^2_{12}}{x_{12}^2}\,,\qquad Y_{12}\equiv Y_1\cdot Y_2\,,\qquad x_{12}\equiv |\vec x_1-\vec x_2|\,.
}
We are interested in studying the four-point function $\langle SSSS\rangle$, which is fixed by conformal and $SU(4)$ symmetry to take the form
 \es{2222}{
 & \langle S(\vec x_1,Y_1) S(\vec x_2,Y_2) S(\vec x_3,Y_3) S(\vec x_4,Y_4) \rangle = \frac{Y^2_{12}Y^2_{34}}{{x}_{12}^4 {x}_{34}^{4}} \mathcal{S}(U,V;\sigma,\tau)\,,
    }
where the conformally invariant cross ratios $U,V$ and the $SO(6)$ invariants $\sigma,\tau$ are
 \es{uvsigmatauDefs}{
  U \equiv \frac{{x}_{12}^2 {x}_{34}^2}{{x}_{13}^2 {x}_{24}^2} \,, \qquad
   V \equiv \frac{{x}_{14}^2 {x}_{23}^2}{{x}_{13}^2 {x}_{24}^2}  \,, \qquad
   \sigma\equiv\frac{(Y_1\cdot Y_3)(Y_2\cdot Y_4)}{(Y_1\cdot Y_2)(Y_3\cdot Y_4)}\,,\qquad \tau\equiv\frac{(Y_1\cdot Y_4)(Y_2\cdot Y_3)}{(Y_1\cdot Y_2)(Y_3\cdot Y_4)} \,.
 }
 Since \eqref{2222} is a degree 2 polynomial in each $Y_i$, the quantity $\mathcal{S}(U,V;\sigma,\tau)$ is a degree 2 polynomial in $\sigma,\tau$. The superconformal Ward identity further requires that $\mathcal{S}(U,V;\sigma,\tau)$ be written in terms of the $SU(4)$-independent quantity $\mathcal{T}(U,V)$ as
 \es{T}{
 \mathcal{S}(U,V;\sigma,\tau)&=\mathcal{S}_\text{free}(U,V;\sigma,\tau)+\Theta(U,V;\sigma,\tau)\mathcal{T}(U,V)\,,\\
 \Theta(U,V;\sigma,\tau)&\equiv\tau+[1-\sigma-\tau]V+\tau[\tau-1-\sigma]U+\sigma[\sigma-1-\tau]UV+\sigma V^2+\sigma\tau U^2\,,
 }
where $\mathcal{S}_\text{free}(U,V;\sigma,\tau)$ is the ``free theory'' correlator\footnote{For a generic $\mathcal{N}=4$ conformal manifold, which need not have a free point, this form is still required by superconformal symmetry.}
\es{free}{
\mathcal{S}_\text{free}(U,V;\sigma,\tau)=1+U^2\sigma^2+\frac{U^2}{V^2}\tau^2+\frac{1}{c}\left(U\sigma+\frac UV\tau+\frac{U^2}{V}\sigma\tau\right)\,.
}
All the non-trivial interacting information in the correlator is contained in $\mathcal{T}(U,V)$, which will be our main focus of study in this paper.

\subsection{Strong coupling expansion and the flat space limit}
\label{strong0}

We now specify to SYM, and discuss the strong coupling 't Hooft limit, where we take $N\to\infty$ (or $c\to\infty$) with $\lambda\equiv g_\text{YM}^2N$ fixed and then $\lambda\to\infty$. Recall from the introduction that the double expansion in $c^{-1}$ and $\lambda^{-\frac12}$ is dual to the IIB expansion in $g_s^2\ell_s^8$ (counting supergraviton loops) and $\ell_s^2$ (counting higher derivatives) according to the AdS/CFT dictionary \eqref{dict}. Higher powers in $c^{-1}$ can thus correspond to either higher genus loop Witten diagrams or corrections to contact Witten diagrams, as we will see below.

It is convenient to express the strong coupling expansion in Mellin space. The Mellin transforms $\mathfrak{M}(s,t;\sigma,\tau)$ and $\cM(s,t)$ of the connected full and reduced correlators $\cS(U,V;\sigma,\tau)_\text{conn}$ and $\cT(U,V)$, respectively, are defined by \cite{Rastelli:2017udc}:
\es{mellinDefD}{
\cS(U,V;\sigma,\tau)_\text{conn.}=&\int_{-i\infty}^{i\infty}\frac{ds\, dt}{(4\pi i)^2} U^{\frac s2}V^{\frac {t}{2}-2} \Gamma^2\left[2-\frac s2\right]\Gamma^{2}\left[2-\frac t2\right]\Gamma^{2}\left[2-\frac {u+4}{2}\right] \mathfrak{M}(s,t;\sigma,\tau)\,,\\
\cT(U,V)=&\int_{-i\infty}^{i\infty}\frac{ds\, dt}{(4\pi i)^2} U^{\frac s2}V^{\frac {t}{2}-2} \Gamma^2\left[2-\frac s2\right]\Gamma^{2}\left[2-\frac t2\right]\Gamma^{2}\left[2-\frac {u}{2}\right] \cM(s,t)\,,\\
}
where $u = 4 - s - t$, and where the integration contours include all poles of the Gamma functions on one side or the other of the contour. The Mellin transform $\mathfrak{M}(s,t;\sigma,\tau)$ of the full correlator is defined such that a bulk contact Witten diagram coming from a vertex with $2m$ derivatives gives rise to a polynomial $\mathfrak{M}(s,t;\sigma,\tau)$ of degree $m$, and similarly an exchange Witten diagrams corresponds to $\mathfrak{M}(s,t;\sigma,\tau)$ with poles for the twists (dimension minus spin) of each exchanged operator. The reduced correlator Mellin amplitude $\cM(s,t)$ is then related to $\mathfrak{M}(s,t;\sigma,\tau)$ by the Mellin space version of  $\Theta(U,V;\sigma,\tau)$ in \eqref{T}, which takes the form of a difference operator given in \cite{Rastelli:2017udc} whose explicit form we will not use. The degree of a given term in $\cM(s,t)$ is four less than that of $\mathfrak{M}(s,t;\sigma,\tau)$ in the large $s,t$ limit due to this difference operator.

The main utility of the Mellin amplitude $\mathcal{M}(s,t)$ for us is that it provides an easy way to relate the holographic correlator $\langle SSSS\rangle$ on $AdS_5\times S^5$ to the IIB S-matrix $\mathcal{A}$ according to the flat space limit formula \cite{Penedones:2010ue,Fitzpatrick:2011hu,Chester:2018aca,Binder:2018yvd,Binder:2019jwn}:
 \es{Gotfst}{
   f(s, t) = \frac{stu}{2048\pi^2g_s^2\ell_s^8 }\lim_{L/\ell_s \to \infty} L^{14} \int_{\kappa-i\infty}^{\kappa+ i \infty} \frac{d\alpha}{2 \pi i} \, e^\alpha \alpha^{-6} {\cal M} \left( \frac{L^2}{2 \alpha} s, \frac{L^2}{2 \alpha} t \right) \,,
 }
 where $f(s,t)$ was defined in \eqref{A} as $\mathcal{A}(s,t)=\mathcal{A}_\text{SG}(s,t)f(s,t)$ so that the leading supergravity term is normalized to one, and the momenta of the flat space S-matrix is here restricted to 5 dimensions. From this flat space formula as well as the AdS/CFT dictionary \eqref{dict} we see that at order $\lambda^{n} c^{-m}$ in the strong coupling expansion, only terms that at large $s$ and $t$ scale as $s^a t^b$ with $a+b=4m-2n-7$ contribute to \eqref{Gotfst}, and have coefficient multiplied by $g_s^{2m-2} \ell_s^{8m-4n-8}$. For instance, the leading supergravity term in the CFT correlator is proportional to $\frac{1}{c}=\frac{4}{N^2-1}$, so in this case $m=1$, $n= 0$, and $a+b=-3$, which corresponds to a constant S-matrix term $f(s,t)$ consistent with our convention. 
 
The Mellin amplitude $\cM(s,t)$ must satisfy two more constraints in addition to the flat space limit. Firstly, it must satisfy the crossing equations
 \es{crossM}{
  \cM(s,t) = \cM(s,u)= \cM(u,t) \,.
 } 
Secondly, large $N$ counting \cite{Aharony:2016dwx} requires that $\cM(s,t)$ receives contributions from exchange Witten diagrams of only single and double-trace operators at tree level, and at most $(n+1)$-trace operators at $n$-loop level, so only poles corresponding to the twists of these operators may appear at each order. These conditions severely restrict the allowed Mellin amplitudes at each order, and lead to the strong coupling expansion shown in \eqref{MellinIntro}, which can then be transformed to position space using \eqref{mellinDefD} to get
\es{TIntro}{
   \cT =&\frac1 c\left[8\cT^\text{SG}+\lambda^{-\frac32}B_0^0 \cT^{0}  +\lambda^{-\frac52}\left[B_{2}^2 \cT^{2}+B_0^2\cT^{0}\right]+O(\lambda^{-3})\right]\\
   &+\frac{1}{c^2}
   \left[\lambda^{\frac12} \overline{B}_0^0 \cT^{0}+\left[\mathcal{T}^{\text{SG}|\text{SG}}+\overline{B}^{\text{SG}|\text{SG}}_0\mathcal{T}^0\right] +\lambda^{-\frac12}\left[ \overline{B}_2^2 \cT^{2} + \overline{B}_0^2 \cT^{0}  \right]+O(\lambda^{-1})\right]\\
   &+\frac{1}{c^3}
   \left[\lambda^{\frac52} \overline{\overline{B}}_0^0 \cT^{0}+\lambda^{\frac32}\left[ \overline{\overline{B}}_2^2 \cT^{2} + \overline{\overline{B}}_0^2 \cT^{0}  \right]+O(\lambda)\right]+O(c^{-4})\,.
}

We will now review the derivation of this expansion in Mellin and position space at each order in $1/c$. At tree level, only single and double-trace operators can be exchanged. The double-trace poles in Mellin space at this order are already taken into account by the Gamma functions in \eqref{mellinDefD}. The only single-trace operators that contribute are those in the supergraviton multiplet with Mellin amplitude $\mathcal{M}^\text{SG}$ in \eqref{MellinIntro}, which takes the simple form \cite{Goncalves:2014ffa,Arutyunov:2000py}
\es{SintM}{
\cM^\text{SG}=&\frac{1}{(s-2)(t-2)(u-2)}\qquad\Rightarrow\qquad \mathcal{T}^\text{SG}= -\frac18U^2\bar D_{2,4,2,2}(U,V)\,,
}
where the position space expression can be found by taking the inverse Mellin transform \eqref{mellinDefD} and is written in terms of the functions $\bar D_{r_1,r_2,r_3,r_3}(U,V)$ defined in \cite{Eden:2000bk}. The coefficient of $\cM^\text{SG}$ is fixed by requiring that the unprotected R-symmetry singlet of dimension two that appears in the conformal block decomposition of the free part ${\cS}_\text{free}(U,V;\sigma,\tau)$ is not present in the full correlator \cite{Rastelli:2017udc}.  In our conventions \cite{Binder:2019jwn}, this amounts to setting the coefficient of $\mathcal{M}^\text{SG}$ to $8/c$. 

At higher order in $1/\lambda$ contact Witten diagrams contribute whose vertices are higher derivative corrections to tree level supergravity of the form $D^{2n}R^4$, which scale as $c^{-1}\lambda^{-\frac {n+3}{2}}$. In Mellin space these terms must be crossing symmetric degree $m$ polynomials $\cM^m$ in $s,t,u$ subject to $s+t+u=4$, where the flat space limit requires that $m\leq n$. The first couple of terms are \cite{Alday:2014tsa}
\es{treePoly}{
\cM^0&=1\qquad\qquad\qquad\;\;\Rightarrow\qquad  \mathcal{T}^0=U^2\bar D_{4,4,4,4}(U,V)\,,\\
 \cM^{2}&=s^2+t^2+u^2 \qquad\Rightarrow\qquad  \mathcal{T}^2= 4U^2\left([1+U+V]\bar D_{5,5,5,5}(U,V)-4\bar D_{4,4,4,4}(U,V)\right)\,,
}
so that at order $c^{-1}\lambda^{-\frac32}$, i.e. $R^4$, only $\cM^0$ contributes with coefficient $B_0^0$, while at order $c^{-1}\lambda^{-\frac52}$, i.e. $D^4R^4$, both $\cM^0$ and $\cM^2$ can contribute with coefficients $B_0^2$ and $B_2^2$, respectively. These coefficients were fixed using localization and the relation to the known IIB S-matrix in the flat space limit \cite{Binder:2019jwn}, and we gave the results in \eqref{oldBs}. Note that at $O(c^{-1})$ only the genus-zero coefficients of these contact Witten diagrams appear, and there could be higher genus terms at higher order in $1/c$, which would still be tree level in the bulk correlator.

At 1-loop, both single and double-trace operators can be exchanged. The single-trace poles were already fixed by the conformal Ward identity to only appear in $\mathcal{M}^\text{SG}$ at order $c^{-1}$, so they do not appear at any other order. The double-trace pole contribution in position space comes from a 1-loop Witten diagram that can be computed by ``squaring'' the contribution of tree level double-trace anomalous dimensions according to the unitarity method of \cite{Aharony:2016dwx}, and so their coefficients are entirely fixed by tree level data. For instance, the leading order in $1/\lambda$ double-trace pole contribution $\mathcal{T}^{\text{SG}|\text{SG}}$ comes from 1-loop Witten diagrams with pairs of supergravity vertices, and so scales as $c^{-2}$. The $\log^2 U$ and $\log^2 V$ terms\footnote{These can be understood as the terms that contribute to the double discontinuity \cite{Caron-Huot:2017vep} at  $O(c^{-2})$.} in $\mathcal{T}^{\text{SG}|\text{SG}}$ were fixed from summing supergravity double-trace anomalous dimensions in \cite{Alday:2017xua}, and in \cite{Aprile:2017bgs} these terms were completed to the full $\mathcal{T}^{\text{SG}|\text{SG}}$ using an ansatz that was verified in \cite{Alday:2017vkk}. The explicit form of $\mathcal{T}^{\text{SG}|\text{SG}}$ is extremely complicated, so we refer the reader to \cite{Aprile:2017bgs} for the explicit definition. The Mellin space expression $\mathcal{M}^{\text{SG}|\text{SG}}$, which was derived in \cite{Alday:2018kkw} from the previous position space expressions, takes the much simpler form
\es{M1loop}{
\mathcal{M}^{\text{SG}|\text{SG}}=&\sum_{m,n=2}\left[\frac{c_{mn}}{5(m+n-5)_5}\left(\frac{1}{(s-2m)(t-2n)}+\frac{1}{(t-2m)(u-2n)}\right.\right.\\
&\left.\left.+\frac{1}{(u-2m)(s-2n)}\right)-d_{mn}\right]+C\,,
}
which has poles at all the expected double-trace twists, and where $c_{mn}$ is
\es{cmn}{
c_{mn}=& 30m^2n^2(m + n)^2 - 10mn(7m^3 + 36m^2n + 36mn^2 + 7n^3) - 296(m + n) + 64\\
&+(44m^4 + 548m^3n + 1152m^2n^2 + 548mn^3 + 44n^4)\\
&-2(128m^3 + 631m^2n + 631mn^2 + 128n^3)+ 12(37m^2 + 90mn + 37n^2)\,.
}
In \cite{Alday:2018kkw}, this expression was given without $d_{mn}$ or $C$, and had a divergence that was independent of $s,t$. We can choose $d_{mn}$ so as to cancel this divergence, and then fix $C$ so that $\mathcal{M}^{\text{SG}|\text{SG}}$ is actually the Mellin transform of $\mathcal{T}^{\text{SG}|\text{SG}}$ as defined in \cite{Aprile:2017bgs}. In Appendix \ref{1loopAp} we show that one (non-unique) choice is
\es{dc}{
d_{mn}=\frac{9mn}{2(m+n)^3}\,,\qquad C=9 \zeta (3)-\frac{39}{16}-\frac{13 \pi ^2}{8}\,.
}
 It was shown in  \cite{Alday:2018kkw} that $\mathcal{M}^{\text{SG}|\text{SG}}$ is asymptotically linear in $s,t$, as expected for a $c^{-2}$ term from the flat space limit, so the full $c^{-2}$ contribution includes a constant $\cM^0$ whose coefficient $B^{\text{SG}|\text{SG}}_0$ cannot be fixed from tree level. Since different choices of $d_{m,n}$ and $C$ can be related by shifting $B^{\text{SG}|\text{SG}}_0$, this coefficient parameterizes the finite counterterm from regulating the divergence \footnote{There is no such ambiguity in flat space, because it is sub-leading in $s,t$ so disappears in the flat space limit.} of 1-loop supergravity on $AdS_5\times S^5$. The next lowest 1-loop term is $\mathcal{M}^{\text{SG}|R^4_\text{genus-0}}$, which scales as $c^{-2}\lambda^{-\frac32}$ and was computed using similar methods in \cite{Alday:2018pdi,Alday:2018kkw}, but we will not consider 1-loop terms at this order.

All 1-loop terms scale as $O(c^{-2})$, but some $O(c^{-2})$ terms are in fact tree level, and can be distinguished from 1-loop terms by their scaling in $\lambda$ for low orders in $1/\lambda$. In particular, the same polynomial Mellin amplitudes, which correspond to tree level contact Witten diagrams, that contributed at $O(c^{-1})$ can also contribute at higher order in $1/c$ if they receive higher genus corrections. For instance, the $R^4$ contact term can receive a genus-one correction that scales as $c^{-2}\lambda^{\frac{1}{2}}$ with Mellin amplitude $\cM^0$ and coefficient $\overline{B}_0^0$, which is in fact more leading than the first 1-loop term $\mathcal{M}^{\text{SG}|\text{SG}}$ that scales as $c^{-2}$. The $D^4R^4$ contact term could also receive a genus-one correction\footnote{From IIB string theory we expect that no such term exists, and we will in fact show that from CFT later.} that scales as $c^{-2}\lambda^{-\frac{1}{2}}$ and can receive contributions from both $\cM^0$ and $\cM^0$ with coefficients $\overline{B}_0^2$ and $\overline{B}_2^2$, respectively. At higher order in $1/\lambda$, which we will not consider in this work, polynomial Mellin amplitudes can also have $\log\lambda$ coefficients that come from regularizing the logarithmic divergences of higher order loop terms such as $\cM^{\text{SG}|R^4_\text{genus-0}}$.\footnote{These logarithmic divergences do not occur for the supergravity-supergravity 1-loop term.} At order $c^{-2}\lambda^{-\frac32}$, we can no longer distinguish between 1-loop and higher genus tree level terms just based on their scaling in $\lambda$, since both the 1-loop $\cM^{\text{SG}|R^4_\text{genus-0}}$ and the genus-one correction to the $D^8R^4$ tree level term contribute at this order. Note that unlike the non-analytic exchange Mellin amplitudes at 1-loop, the tree level polynomial contact Mellin amplitude have coefficients $\overline{B}_0^0$, $\overline{B}_2^2$, and $\overline{B}_0^2$ that cannot be fixed from $O(c^{-1})$ data.

The story at higher $1/c$ is similar to $O(c^{-2})$. There will be non-analytic loop terms that can in principle be fixed from lower loop order, and polynomial tree level terms with unfixed coefficients. At $O(c^{-3})$, the leading order loop term is the 1-loop $\mathcal{M}^{\text{SG}|R^4_\text{genus-one}}$, and so scales as $c^{-3}\lambda^{\frac12}$. This term is subleading to the two leading order tree level polynomial Mellin amplitudes: genus-two\footnote{This term, and in fact all higher genus $R^4$ terms, will later be shown to vanish.} $R^4$ that scales as $c^{-3}\lambda^{\frac52}$ and includes $\cM^0$ with coefficient $\overline{\overline{B}}_0^0$, and genus-two $D^4R^4$ that scales as $c^{-3}\lambda^{\frac32}$ and includes $\cM^0$ and $\cM^2$ with coefficients $\overline{\overline{B}}_0^2$ and $\overline{\overline{B}}_2^2$, respectively. These are the highest order terms we will consider in this work.

Our goal is now to fix all of the $\cM^0$ coefficients $\overline{B}_0^0$, $\overline{B}_0^{\text{SG}|\text{SG}}$, $\overline{B}_0^2$, $\overline{\overline{B}}_0^0$, $\overline{\overline{B}}_0^2$ and all the $\cM^2$ coefficients $\overline{B}_2^2$, $\overline{\overline{B}}_2^2$ that appear in \eqref{MellinIntro} (equivalently \eqref{TIntro}). To do this we will use the relation to the known IIB S-matrix using the flat space limit formula \eqref{Gotfst}, as well as the relation to the $\mathcal{N}=2^*$ free energy on $S^4$ that was shown in \cite{Binder:2019jwn}, which we will review next.

\subsection{Relation to $\mathcal{N}=2^*$ free energy on $S^4$}
\label{strong}

The action of any 4d $\mathcal{N}=4$ SCFT can be deformed by the complex marginal operator $\Phi$ that couples to the complex conformal manifold parameter $\tau$. We can also deform by the dimension two scalar $S$ and the dimension three scalar $P$ in the stress tensor multiplet, which both couple to a real mass $m$ and break the supersymmetry to $\mathcal{N}=2$. Since correlators of all the operators in the stress tensor multiplet are related by supersymmetry \cite{Dolan:2001tt,Belitsky:2014zha}, the parameters $m$, $\tau$, and $\bar\tau$ are sources for all stress tensor multiplet correlators. In \cite{Binder:2019jwn}, it was shown that the correlator $\langle SSSS\rangle$ integrated over $S^4$ was related to the $S^4$ free energy $F(m,\tau,\bar\tau)$ as
 \es{constraint1}{
c^2I[\cT(U,V)]=&\frac{c}{8}\frac{\partial_m^2\partial_\tau\partial_{\bar\tau}F}{\partial_\tau\partial_{\bar\tau}F}\Big\vert_{m=0}\,,\\
}
where $\cT$ is the interacting part of $\langle SSSS\rangle$ as defined in \eqref{T}, and $I[\mathcal{G}]$ is the $S^4$ integral
\es{I}{
I[\cG(U,V)]\equiv&\frac{4}{ \pi}  \int dr\,  d\theta \, r^3 \sin^2 \theta \, 
	\frac{r^2 - 1 - 2 r^2 \log r}{(r^2 - 1)^2} \frac{  {\cG}\left(1 + r^2 - 2 r \cos \theta , r^2 \right)  }{ (1 + r^2 - 2 r \cos \theta)^{2} } \,.
 }

As shown by Pestun \cite{Pestun:2007rz}, the $S^4$ partition function $Z=\exp(-F)$ of mass deformed $\mathcal{N}=4$ SYM, i.e. the ${\cal N} = 2^*$ theory, with gauge group $SU(N)$ can be computed using supersymmetric localization through a matrix model that takes the form
\es{N2starMatrixModel}{
	Z(m, \lambda)
	= \int d^{N} a\,\delta\big(\sum_i a_i\big) e^{-\frac{8 \pi^2 N }{\lambda} \sum_i a_i^2} \abs{Z_\text{inst}}^2  \frac{ \prod_{i < j}a_{ij}^2 H^2(a_{ij})}{ H(m)^{N-1} \prod_{i \neq j} H(a_{ij}+ m)} \,,
}
where $a_{ij}\equiv a_{ij}$, the delta function enforces that the $SU(N)$ eigenvalues $a_i$ have zero trace, we define $\lambda\equiv \frac{4\pi N}{\Im\tau}$, and $H(z) = e^{-(1+\gamma)z^2}G(1+ i z) G(1-i z)$ is a product of two Barnes G-functions.  The quantity $\abs{Z_\text{inst}}^2$ represents the contribution to the localized partition function coming from instantons located at the North and South poles of $S^4$ \cite{Nekrasov:2002qd,Nekrasov:2003rj,Losev:1997tp,Moore:1997dj}, and can be ignored in the 't Hooft limit because it is exponentially small when $g_\text{YM} \to 0$. At $m=0$, the partition function describes a Gaussian matrix model, whose free energy takes the form \cite{Pestun:2007rz}
\es{free1}{
F(0,\lambda)=-2c\log\lambda+\text{$\lambda$-independent term}\,,
}
so that $\partial_\tau\partial_{\bar\tau}F=-\frac{c\lambda^2}{32\pi^2N^2}$. The RHS of the integrated constraint \eqref{constraint1} can then be simplified for $SU(N)$ SYM in the 't Hooft limit to 
\es{simpInt}{
\cF\equiv -\frac{1}{16\lambda^2}{\partial_m^2\partial^2_{\lambda^{-1}}F}\big\vert^\text{pert}_{m=0}\,,
}
where $F^\text{pert}$ denotes the perturbative free energy that ignores instantons.
This quantity was computed to leading order in $1/N^2$ in the 't Hooft limit in \cite{Russo:2013kea} using a large $N$ saddle point approximation to get
\es{fFRusso}{
\cF=N^2\int_{0}^\infty d\omega\, \omega \frac{J_1(\frac{\sqrt{\lambda}}{\pi}\omega)^2-J_2(\frac{\sqrt{\lambda}}{\pi}\omega)^2}{4\sinh^2\omega}+O(N^0)\,.
}
 It was then expanded to any order in $1/\lambda$ in \cite{Binder:2019jwn} to get
 \es{fFRusso2}{
\frac{\cF}{c^2}=\frac{1}{c} \left(\frac14-\frac{3\zeta(3)}{\lambda^{\frac32}}+\frac{45\zeta(5)}{4\lambda^{\frac52}}+\dots\right)+ O(c^{-2})\,,
}
where we converted from $1/N^2$ to $1/c$ using $c=\frac{N^2-1}{4}$, and divided by $c^2$ to take into account that factor on the LHS of \eqref{constraint1}. This $O(c^{-1})$ expression along with the integrated constraint \eqref{constraint1} and the flat space limit was used in \cite{Binder:2019jwn} to fix the $O(c^{-1})$ coefficients $B_0^0$, $B_0^2$, and $B_2^2$ in \eqref{MellinIntro}. In the next section, we will generalize \eqref{fFRusso} to all orders in $1/N^2$ and $1/\lambda$, which can then be used to fix the remaining coefficients shown in \eqref{MellinIntro}.

\section{$\mathcal{N}=2^*$ free energy on $S^4$}
\label{freem}

The goal of this section is to compute the quantity $\cF$ in \eqref{simpInt} to all orders in $1/N^2$ and $1/\lambda$ using the $\mathcal{N}=2^*$ $SU(N)$ free energy $F(m,\lambda)$ on $S^4$, where we can ignore the contribution from instantons in the 't Hooft limit. From the localized partition function \eqref{N2starMatrixModel}, we see that $\cF\sim \partial_m^2 F\big\vert^\text{pert}_{m=0}$ can be expressed as a matrix model expectation value of a 2-body operator
 \es{2m2L}{
\cF=\frac{1}{16\lambda^2}\partial_{\lambda^{-1}}^2{\sum_{i, j}\langle K'(a_{ij})\rangle}\,,
 }
 where $K(z)\equiv -\frac{H'(z)}{H(z)}$, and $K'(z)$ can be simply expressed using its Fourier transform
  \es{Kfourier}{
 K'(z)=-\int_0^\infty d\omega\frac{2\omega[\cos(2\omega z)-1]}{\sinh^2\omega}\,.
 }
The expectation value should be taken with respect to the matrix model of the $m=0$ theory, i.e. $\mathcal{N}=4$ $SU(N)$ SYM. Since the operator $K'(a_{ij})$ only depends on the difference between eigenvalues, its expectation value is in fact the same for $SU(N)$ or $U(N)$ SYM, so for simplicity we will consider the $U(N)$ partition function
\es{Zfree}{
Z^{U(N)}(0, \lambda)
	= \int d^{N} a\, e^{-\frac{8 \pi^2 N }{\lambda} \sum_i a_i^2}  \prod_{i < j} a_{ij}^2\,.
	}
We can also ignore the $-1$ term in \eqref{Kfourier}, since we take derivatives of $\lambda$ in \eqref{2m2L}. Our goal is then to compute the expectation value 
\es{cos}{
\sum_{i, j}\langle \cos(2\omega(a_{ij}))\rangle=\sum_{i, j}\langle e^{2i\omega(a_{ij})}\rangle\,,
}
where we used the fact that the sum is symmetric in $i,j$. This expectation value is very similar to that of $\mathcal{N}=4$ Wilson loops, which have been computed in an $1/N^2$ expansion for finite $\lambda$ using topological recursion \cite{Akemann:2001st,Okuyama:2017feo,Okuyama:2018aij}, and also for finite $N$ and $\lambda$ using orthogonal polynomials \cite{Drukker:2000rr,Fiol:2013hna}. We will now apply the same methods to \eqref{cos}, and then take the integral over the auxiliary variable $\omega$ in \eqref{Kfourier} and take the $\lambda$ derivatives in \eqref{2m2L} to recover $\cF$. 

\subsection{$1/N^2$ expansion from topological recursion}
\label{top}

The strategy of this calculation is to express the 2-body expectation value \eqref{cos} in terms of a quantity called the resolvent, which has a known expansion to all orders in $1/N$. We do this by expressing  \eqref{cos} in terms its inverse Laplace transform with respect to each argument:
\es{expToW}{
\sum_{i, j}\langle e^{2i\omega(a_{ij})}\rangle=& N^2\cL^{-1}[W^1(y_1)](2i\omega)\;\cL^{-1}[W^1(y_2)](-2i\omega)+\cL^{-1}[W^2(y_1,y_2)](2i\omega,-2i\omega)\,,\\
}
where the inverse Laplace transform is defined as
\es{L}{
&\cL^{-1}[f(y_1,\dots,y_n)](b_1,\dots, b_n)\equiv \frac{1} {(2\pi i)^n} \left[\prod_{i=1}^n\int_{\gamma_i-i\infty}^{\gamma_i+i\infty}dy_i e^{b_i y_i} \right]f(y_1,\dots,y_n)\,,
}
with $\gamma_{i}$ chosen so that the contour lies to the right of all singularities in the integrand, and we have included the $i=j$ term in \eqref{expToW}, which is independent of $\lambda$ and so does not affect $\cF$. The resolvent $W(y_1,\dots, y_n)$ is defined as the connected expectation value
\es{W}{
W^n(y_1,\dots, y_n)\equiv N^{n-2}\big\langle\sum_{i_1} \frac{1}{y_1 -a_{i_1} }\cdots \sum_{i_n} \frac{1}{y_n -a_{i_n} }\big\rangle_\text{conn.}\,,
}
which has the large $N$ expansion 
\es{W2}{
W^n(y_1,\dots, y_n)\equiv\sum_{m=0}^\infty\frac{1}{N^{2m}} W^n_m(y_1,\dots, y_n)\,.
}
The coefficients $W^n_m$ are generating functions of ``genus'' $m$ discrete surfaces with $n$ boundaries, so this expansion is called the ``genus'' expansion \cite{DiFrancesco:1993cyw}.\footnote{This ``genus'' expansion is just the $1/N^2$ expansion, which differs from the $\frac{1}{c}=\frac{4}{N^2-1}$ expansion that counts higher genus corrections in the holographic correlator. To avoid confusion between the two uses of genus, we will refer to ``genus'' in the resolvent expansion using quotes.} These $W^n_m$ can be computed for any $n,m$ in a Gaussian matrix model using the topological recursion method of \cite{Eynard:2004mh,Eynard:2008we}. We start with the ``genus'' zero  1-body resolvent
\es{W01}{
W^1_0(y_1)=\frac1\lambda \left({8 \pi ^2 y_1}-{8 \pi ^2 \sqrt{y_1^2-\frac{\lambda }{4
   \pi ^2}}}\right)\,.
}
The other ``genus'' zero resolvents can be computed from the recursion formula
\es{Rec1}{
W^n_0(y_1,\dots ,y_n)=& \frac{\lambda}{ 16 \pi ^2\sqrt{y_1^2-\frac{\lambda }{4 \pi ^2}}}  \Bigg[\sum_{l=1}^{n-2}\sum_{I\in R^n_l}W_0^{l+1}(y_1,y_I)W_0^{n-l}(y_1,y_{R^n-I})\\
&+\sum_{l=2}^n\partial_{y_l}\frac{W_0^{n-1}(y_2,\dots,y_l,\dots,y_n)-W_0^{n-1}(y_2,\dots,y_1,\dots,y_n)}{y_l-y_1}\Bigg]\,,
}
where $R^n=\{2,\dots,n\}$ and $R_l^n$ are subsets of $R^n$ of size $l$. The higher ``genus'' resolvents can then be computed from the recursion formulae for $m\geq 1$:
\es{Rec2}{
W^1_m(y_1)=& \frac{\lambda}{ 16 \pi ^2\sqrt{y_1^2-\frac{\lambda }{4 \pi ^2}}}  \left[W^2_{m-1}(y_1,y_1)+\sum_{r=1}^{m-1}W^1_{m-r}(y_1)W^1_r(y_1)\right]\,,
}
and for $m\geq 1$ and $n\geq2$:
\es{Rec3}{
W^n_m(y_1,\dots ,y_n)=& \frac{\lambda}{ 16 \pi ^2\sqrt{y_1^2-\frac{\lambda }{4 \pi ^2}}}  \Bigg[W^{n+1}_{m-1}(y_1,y_1,\dots y_n)+2\sum_{r=1}^{m-1}W^1_{m-r}(y_1)W^1_r(y_1)\\
&+
\sum_{r=0}^m\sum_{l=1}^{n-2}\sum_{I\in R^n_l}W_r^{l+1}(y_1,y_I)W_{m-r}^{n-l}(y_1,y_{R^n-I})\\
&+\sum_{l=2}^n\partial_{y_l}\frac{W_m^{n-1}(y_2,\dots,y_l,\dots,y_n)-W_m^{n-1}(y_2,\dots,y_1,\dots,y_n)}{y_l-y_1}\Bigg]\,.
}
These last two recursion formulae compute all $W^n_m$ in terms of $W^{n'}_{m'}$ with $n'+m'<n+m$. 

The recursion formulae can be used to efficiently compute resolvents to any order. For instance, the ``genus'' zero 2-body resolvent is \cite{Ambjorn:1990ji}
\es{W20}{
W^2_0(y_1,y_2)=\frac{\frac{4 \pi ^2 y_1 y_2}{\lambda }-1- \sqrt{\frac{4 \pi ^2
   y_1^2}{\lambda }-1} \sqrt{\frac{4 \pi ^2 y_2^2}{\lambda }-1}}{2
   \left(y_1-y_2\right){}^2 \sqrt{\frac{4 \pi ^2 y_1^2}{\lambda }-1}
   \sqrt{\frac{4 \pi ^2 y_2^2}{\lambda }-1}}\,.
}
In Appendix \ref{res} we give the other resolvents we need to compute \eqref{expToW} to $O(N^{-6})$, i.e. $W^1_m$ and $W^2_{m-1}$ for $m=1,\dots,4$. In general, all $W^n_m$ factor in terms of their arguments $y_i$, except for $W^2_0$. For the former resolvents, it is straightforward to take the inverse Laplace transforms for each $y_i$ separately, which yield a Bessel function for each $y_i$. For instance, the inverse Laplace transform of $W_0^1$ in \eqref{W01} with argument $2i\omega$ is
\es{W01Bessel}{
 \cL^{-1}[ W^1_0(y_1)](2i\omega)=\frac{2\pi J_1(\frac{\sqrt{\lambda } \omega }{\pi })}{\omega\sqrt{\lambda}} \,,
}
and we give the results for the other factorizable resolvents in Appendix \ref{res}. For $W^2_0(y_1,y_2)$, we must perform the two-dimensional integral in \eqref{L} with the specific arguments $b_1=-b_2=2\omega i$ to get\footnote{The inverse Laplace transform for all arguments $b_1\neq -b_2$ was given in \cite{Akemann:2001st}, but that result is singular if we naively set $b_1=-b_2$ in that formula.}
\es{W20int}{
& \cL^{-1}[ W^2_0(y_1,y_2)](2i\omega,-2i\omega)=\frac{\lambda  \omega ^2}{2\pi^2}\left[  \textstyle{ J_0({\frac{\sqrt{\lambda } \omega }{\pi
   }})^2}+
{   J_1(\frac{\sqrt{\lambda } \omega }{\pi })^2}\displaystyle-\frac{ \pi J_1(\frac{\sqrt{\lambda } \omega
   }{\pi }) J_0(\frac{\sqrt{\lambda } \omega }{\pi })}{\lambda\omega}\right]\,.
}

Now that we can compute \eqref{expToW} to arbitrary order in $1/N^2$ in terms of products of two Bessel functions, we can then use \eqref{2m2L}, \eqref{Kfourier}, and \eqref{cos} to compute $\cF$ as an expansion in $1/N^2$ as
\es{Fexp}{
\cF\equiv\sum_{m=0}^\infty \frac{1}{N^{2(m-1)}}\widetilde\cF_m\,,
}
where we included the overall $N^2$ in \eqref{expToW} and 
\es{FtildeGen}{
\widetilde\cF_m=&-\frac{1}{8\lambda^2}\int_0^\infty d\omega \frac{\omega}{\sinh^2\omega}\partial_{\lambda^{-1}}^2\Bigg[\cL^{-1}[W_{m-1}^2(y_1,y_2)](2i\omega,-2i\omega)] \\
&+\sum_{r=0}^m \cL^{-1}[W_r^1(y_1)](2i\omega)\;\cL^{-1}[W_{m-r}^1(y_2)](-2i\omega)\Bigg]\,.
}
For $\widetilde\cF_0$ we ignore the first term in \eqref{FtildeGen} and use the inverse Laplace transform of $W^1_0$ in \eqref{W01Bessel} to get 
\es{Ftilde}{
\widetilde\cF_0=\int_{0}^\infty d\omega\, \omega \frac{J_1(\frac{\sqrt{\lambda}}{\pi}\omega)^2-J_2(\frac{\sqrt{\lambda}}{\pi}\omega)^2}{4\sinh^2\omega}\,,
}
which matches the expression \eqref{fFRusso} originally computed in \cite{Russo:2013kea} using a large $N$ saddle point approximation. For $m=1,2,3,4$, we use the explicit inverse Laplace transforms of the resolvents in Appendix \ref{res} and \eqref{W20int} to find
\es{higherW}{
\widetilde\cF_1=&\int_0^\infty d\omega\frac{-\lambda\omega^3}{2^63\pi^3\sinh^2\omega}\left[\textstyle2{\sqrt{\lambda}\omega J_0(\frac{\sqrt{\lambda } \omega }{\pi
   }) J_1(\frac{\sqrt{\lambda } \omega }{\pi })}+ 12\pi
{   J_0(\frac{\sqrt{\lambda } \omega }{\pi })^2}
  +{5\pi  
   J_1(\frac{\sqrt{\lambda } \omega }{\pi })^2}\right]\,,\\
  \widetilde\cF_2=&\int_0^\infty d\omega\frac{\lambda\omega^3}{2^{11}45\pi^6\sinh^2\omega}\Big[  
  \textstyle(-5 \lambda ^2 \omega ^4+230 \pi ^2 \lambda  \omega ^2-48 \pi ^4)
   J_1(\frac{\sqrt{\lambda } \omega }{\pi })^2\\
   &\textstyle+\lambda  \omega
   ^2 \left(5 \lambda  \omega ^2+252 \pi ^2\right)
   J_0(\frac{\sqrt{\lambda } \omega }{\pi })^2-\pi 
   \sqrt{\lambda } \omega  \left(59 \lambda  \omega ^2+480 \pi ^2\right)
   J_1(\frac{\sqrt{\lambda } \omega }{\pi })
   J_0(\frac{\sqrt{\lambda } \omega }{\pi })\Big]\,,\\
 \widetilde\cF_3=&\int_0^\infty d\omega\frac{\lambda  \omega ^3}{2^{16}2835 \pi ^9\sinh^2\omega}\Big[\textstyle3 \pi  \lambda  \omega^2 \left(203 \lambda ^2 \omega^4+4496 \pi ^2 \lambda  \omega^2-19968
   \pi ^4\right) J_0(\frac{\omega \sqrt{\lambda }}{\pi })^2\\
   &\textstyle+4
   \sqrt{\lambda } \omega \left(35 \lambda ^3 \omega^6-537 \pi ^2 \lambda ^2 \omega^4-9816
   \pi ^4 \lambda  \omega^2+24192 \pi ^6\right) J_1(\frac{\omega \sqrt{\lambda
   }}{\pi }) J_0(\frac{\omega \sqrt{\lambda }}{\pi })\\
   &\textstyle+\left(-679
   \pi  \lambda ^3 \omega^6+7788 \pi ^3 \lambda ^2 \omega^4-11136 \pi ^5 \lambda 
   \omega^2+46080 \pi ^7\right) J_1(\frac{\omega \sqrt{\lambda }}{\pi
   })^2\Big]\,,\\
 \widetilde\cF_4=&\int_0^\infty d\omega\frac{\lambda\omega^3}{2^{22}42525\pi^{12}\sinh^2\omega}\Big[\textstyle2 \lambda  \omega ^2 (-175 \lambda ^4 \omega ^8+8934 \pi ^2 \lambda ^3
   \omega ^6+280872 \pi ^4 \lambda ^2 \omega ^4\\
   &\textstyle-4387968 \pi ^6 \lambda 
   \omega ^2+16865280 \pi ^8) J_0(\frac{\sqrt{\lambda } \omega
   }{\pi })^2
-4 \pi  \sqrt{\lambda } \omega  (-2345 \lambda ^4
   \omega ^8\\
   &\textstyle+11718 \pi ^2 \lambda ^3 \omega ^6
   +656496 \pi ^4 \lambda ^2
   \omega ^4-5907456 \pi ^6 \lambda  \omega ^2\\
   &\textstyle+11059200 \pi ^8)
   J_1(\frac{\sqrt{\lambda } \omega }{\pi })
   J_0(\frac{\sqrt{\lambda } \omega }{\pi })+(350 \lambda
   ^5 \omega ^{10}-22733 \pi ^2 \lambda ^4 \omega ^8+239856 \pi ^4 \lambda
   ^3 \omega ^6\\
   &\textstyle-1027008 \pi ^6 \lambda ^2 \omega ^4+10515456 \pi ^8 \lambda
    \omega ^2-46448640 \pi ^{10}) J_1(\frac{\sqrt{\lambda }
   \omega }{\pi })^2\Big]\,.\\
}

These expressions can be expanded to any order in $1/\lambda$ using the Mellin-Barnes formula for a product of Bessel functions as described in Appendix D\footnote{This method of expansion was pointed out to the authors of that paper by \verb|MathOverflow| user \verb|Paul Enta| in \verb|https://mathoverflow.net/questions/315264/asymptotic-expansion-of-bessel-function-integral|.} of \cite{Binder:2019jwn}, which gives
\es{largeLam}{
\widetilde\cF_0=&\frac{1}{16}-\frac{3 \zeta (3)}{4 \lambda ^{3/2}}+\frac{45 \zeta (5)}{16 \lambda
   ^{5/2}}+\frac{4725 \zeta (7)}{512 \lambda ^{7/2}}+\dots\,,\\
   \widetilde\cF_1=&-\frac{\sqrt{\lambda }}{64}-\frac{39 \zeta (3)}{2048 \lambda
   ^{3/2}}-\frac{1125 \zeta (5)}{4096 \lambda ^{5/2}}-\frac{2811375 \zeta
   (7)}{524288 \lambda ^{7/2}}+\dots\,,\\
   \widetilde\cF_2=&\frac{\lambda ^{3/2}}{6144}-\frac{13 \sqrt{\lambda }}{32768}+\frac{4599
   \zeta (3)}{1048576 \lambda ^{3/2}}+\frac{1548855 \zeta (5)}{8388608
   \lambda ^{5/2}}+\frac{2029052025 \zeta (7)}{268435456 \lambda ^{7/2}}+\dots\,,\\
   \widetilde\cF_3=&\frac{\lambda ^{5/2}}{393216}-\frac{25 \lambda ^{3/2}}{1572864}+\frac{1533
   \sqrt{\lambda }}{16777216}-\frac{3611751 \zeta (3)}{1073741824 \lambda
   ^{3/2}}\\
   &-\frac{581627475 \zeta (5)}{2147483648 \lambda
   ^{5/2}}-\frac{2517203563875 \zeta (7)}{137438953472 \lambda ^{7/2}}+\dots\,,\\
   \widetilde\cF_4=&\frac{3 \lambda ^{7/2}}{20971520}-\frac{595 \lambda
   ^{5/2}}{402653184}+\frac{11473 \lambda ^{3/2}}{1073741824}-\frac{1203917
   \sqrt{\lambda }}{17179869184}+\frac{5635016673 \zeta (3)}{1099511627776
   \lambda ^{3/2}}\\
   &+\frac{2936119026555 \zeta (5)}{4398046511104 \lambda
   ^{5/2}}+\frac{9357342327107775 \zeta (7)}{140737488355328 \lambda
   ^{7/2}}+\dots\,.\\
}
Note that the $\lambda^{-\frac12}$ term vanishes for all $\cF_m$, the $\lambda^{-\frac n2}$ terms for positive odd $n>1$ have coefficient $\zeta(n)$, and a constant term only shows up in $\widetilde\cF_0$.

The ``genus'' expansion for the expectation value \eqref{cos} is naturally an expansion in $1/N^2$, since this quantity is the same for $U(N)$ or $SU(N)$ SYM, but for $SU(N)$ SYM we are interested in the $\frac{1}{c}=\frac{4}{N^2-1}$ expansion:
\es{cExp}{
\cF\equiv\sum_{m=0}^\infty\frac{1}{c^{m-1}}\cF_m\,.
}
We can convert the $1/N^2$ expansion coefficients $\widetilde\cF_m$ in \eqref{Fexp} into the $1/c$ expansion coefficients $\cF_m$ as
\es{cToN}{
 \cF_0=4\widetilde\cF_0\,,\qquad
\cF_1=\widetilde\cF_0+\widetilde\cF_1\,,\qquad
\cF_{m\geq2}=\sum_{n=0}^m \frac{(-1)^{m+n}\widetilde\cF_{n+2}}{4^{m-1}}\begin{pmatrix} m-2\\n\end{pmatrix}\,,
}
so that $\cF_0$ matches the $O(c^{-1})$ (i.e genus-zero) term in \eqref{fFRusso2}, both $O(c^{-1})$ (genus-zero) $\cF_0$ and $O(c^{-2})$ (genus-one) $\cF_0$ contain a constant term in the $1/\lambda$ expansion from $\widetilde\cF_0$ in \eqref{largeLam}, and the $O(c^{-3})$ (genus-two) term is simply $\cF_2=\widetilde\cF_2/4$.

\subsection{Finite $N$ from orthogonal polynomials}
\label{ortho2}

Instead of expanding $\cF$ in a 't Hooft expansion for large $N$ (or large $c$), one can compute it for finite $N$ in terms of a single finite sum using the method of orthogonal polynomials \cite{mehta1981}. While this finite $N$ answer is not the full answer for the mass deformed free energy, since we neglected instantons in the definition \eqref{simpInt} of $\cF$, it can still serve as a nontrivial check on the 't Hooft expansion of the previous section.

We begin with the 2-body expectation value \eqref{cos}, which we can write as
\es{cos2}{
\sum_{i, j}\langle e^{2i\omega(a_{ij})}\rangle=\frac{N(N-1)}{2}\langle e^{2i\omega(a_1-a_2)}+e^{2i\omega(a_2-a_1)}\rangle+N\,,
}
since the expectation value is the same for each pair of eigenvalues $a_i\neq a_j$. We now introduce a family of polynomials $p_n(a)$ using the Hermite polynomials $H_n(x)$:
 \es{pn}{
 p_n(a)\equiv \left(\frac{\lambda}{32\pi^2 N}\right)^{\frac n2}H_n\left(\frac{4\pi \sqrt{N}a}{\sqrt{2\lambda}}\right)\,,
 }
 which are orthogonal with respect to the Gaussian measure 
 \es{ortho}{
 \int da\, p_m(a)p_n(a)e^{-\frac{8\pi^2N}{\lambda}a^2}=n!\left(\frac{\lambda}{16\pi^2 N}\right)^n\sqrt{\frac{\lambda}{8\pi N}}\delta_{mn}\equiv h_n\delta_{mn}\,.
 }
These orthogonal polynomials are useful because we can substitute the Vandermonde determinant in the Gaussian matrix model \eqref{Zfree} by a determinant of these polynomials as
\es{VantoOrtho}{
\prod_{i<j}{(a_{ij})^2}=&\prod_{i<j}|p_{i-1}(a_j)|^2\\
=& \sum_{\sigma_1\in S_N}(-1)^{|\sigma_1|}\prod_{k_1=1}^N p_{\sigma_1(k_1)-1}(a_{k_1})\sum_{\sigma_2\in S_N}(-1)^{|\sigma_2|}\prod_{k_2=1}^Np_{\sigma_2(k_2)-1}(a_{k_2})\,,
}
where we expanded the determinant in terms of permutations of its matrix elements. We can now perform each $a_i$ integral in \eqref{Zfree} using the orthogonality relation \eqref{ortho} to get
\es{gauss}{
Z(0,\lambda)=&N!\prod_{k=0}^{N-1}h_k\,.
}

 Let us now consider an $n$-body operator $\cO_n(a)$ that without loss of generality only depends on $a_i$ for $i=1,\dots n$. We can write this expectation value using \eqref{VantoOrtho} as
 \es{orthoExp}{
 \langle\cO_n(a)\rangle=&\frac{1}{Z(0,\lambda)} \int d^{N} a\, \cO_n(a) e^{-\frac{8 \pi^2 N }{\lambda} \sum_i a_i^2} \\
 &\times\sum_{\sigma_1\in S_N}(-1)^{|\sigma_1|}\prod_{k_1=1}^N p_{\sigma_1(k_1)-1}(a_{k_1})\sum_{\sigma_2\in S_N}(-1)^{|\sigma_2|}\prod_{k_2=1}^Np_{\sigma_2(k_2)-1}(a_{k_2})\,.
 }
Due to the orthogonality relation \eqref{ortho}, the only permutations $\sigma_1,\sigma_2$ that survive integration are those for which $\sigma_2(m)=\sigma_1(m)$ for $m>n$. This means that in order to contribute to the full matrix model integral, $\{\sigma_2(1),\dots,\sigma_2(n)\}$ must be a permutation of $\{\sigma_1(1),\dots,\sigma_1(n)\}$, which we denote by $\mu$. The expectation value is then 
 \es{nbodyExp}{
 \langle\cO_n(a)\rangle=\frac{1}{N!}\sum_{\sigma\in S_N}\sum_{\mu\in S_n}(-1)^{|\mu|}\int\left(\prod_{i=1}^n da_i\frac{p_{\sigma(i)-1}(a_i)p_{\mu(\sigma(i))-1}(a_i)}{h_{\sigma(i)-1}}e^{-\frac{8\pi^2N}{\lambda}a_i^2}\right)\cO_n(a)\,,
 }
where we used the expression \eqref{gauss} of $Z(0,\lambda)$ in terms of $h_k$. The originally $N$-dimensional integral has now reduced to an $n$-dimensional integral. For the 2-body operator in \eqref{cos2}, we can perform the integrals in \eqref{nbodyExp} using the identity
 \es{identity}{
 \int_{-\infty}^\infty e^{-x^2+yx}H_m(x)H_n(x)=e^{\frac{y^2}{4}}2^m\sqrt{\pi}m!y^{n-m}L_m^{n-m}(-y^2/2)
 }
 to get the expectation value
 \es{firstExp}{
\sum_{i, j}\langle e^{2i\omega(a_{ij})}\rangle=&\frac{N(N-1)}{N!}e^{\frac{-\omega^2\lambda}{4\pi^2 N}}\sum_{\sigma\in S_N}\left[L_{\sigma(1)-1}\left(\frac{\omega^2\lambda}{4\pi^2N}\right)L_{\sigma(2)-1}\left(\frac{\omega^2\lambda}{4\pi^2N}\right)\right.\\
&\left.\qquad\qquad-(-1)^{\sigma(1)-\sigma(2)}L_{\sigma(1)-1}^{\sigma(2)-\sigma(1)}\left(\frac{\omega^2\lambda}{4\pi^2N}\right)L_{\sigma(2)-1}^{\sigma(1)-\sigma(2)}\left(\frac{\omega^2\lambda}{4\pi^2N}\right)\right]+N\\
=&e^{\frac{-\omega^2\lambda}{4\pi^2 N}}\left[\left[L_{N-1}^1\left(\frac{\omega^2\lambda}{4\pi^2N}\right)\right]^2-\sum_{i,j=1}^N(-1)^{i-j}L_{i-1}^{j-i}\left(\frac{\omega^2\lambda}{4\pi^2N}\right)L_{j-1}^{i-j}\left(\frac{\omega^2\lambda}{4\pi^2N}\right)\right]+N\,,\\
 }
 where $L_a^b(x)$ are generalized Laguerre polynomials. This sum can be easily performed for any $N$ to get a polynomial in $\omega$ times $e^{\frac{-\omega^2\lambda}{4\pi^2 N}}$. We can then use \eqref{2m2L}, \eqref{Kfourier}, and \eqref{cos} to compute $\cF$ for finite $N$ as an integral over this sum. For instance, for $N=2$ we find
 \es{N2}{
 \sum_{i, j}\langle e^{2i\omega(a_{ij})}\rangle\big\vert_{N=2}=e^{-\frac{\lambda  \omega ^2}{8 \pi ^2}} \left(2-\frac{\omega^2\lambda}{2\pi^2 }\right)+2\,,
 }
and then $\cF$ is computed as
 \es{final1}{
  \cF\vert_{N=2}=&\int_0^\infty d\omega\frac{\lambda \omega ^3 e^{-\frac{\lambda  \omega ^2}{8 \pi ^2}}  }{1024 \pi
   ^6\sinh^2\omega}\left(\lambda ^2
   \omega ^4-36 \pi ^2 \lambda  \omega ^2+192 \pi ^4\right)\,,
 }
which can be evaluated numerically for any $\lambda$. We compare the orthogonal polynomial results for $\cF$ to the $O(N^{-6})$ expansion of this quantity in \eqref{Ftilde} and \eqref{higherW} for $N=2,\dots,10$ and several values of $\lambda$, where in both cases we computed the $\omega$ integral numerically for each $\lambda$. The $1/N^2$ expansion is most accurate for small $g^2_\text{YM}=\lambda/N$ and large $N$, but appears to be very precise for all range of parameters. Note that each subsequent $1/N^2$ correction in \eqref{higherW} improves the match to the finite $N$ result.

 \begin{table}
\begin{center}
\begin{tabular}{c||c|c|c|c|c|c|c|c|c}
 $\abs{\frac{\cF^{1/N}-\cF^\text{finite}}{\cF^{1/N}+\cF^\text{finite}}}$   & $N= 2$  & 3 & 4& 5& 6& 7&8&9&10  \\
  \hline
  \hline
  $\lambda= 1/4$ &$10^{-13}$  & $10^{-13}$ & $10^{-13}$ & $10^{-13}$&$10^{-14}$&$10^{-13}$ &$10^{-14}$&$10^{-13}$&$10^{-13}$ \\
  \hline 
   $\qquad 1/2$ & $10^{-14}$ & $10^{-14}$  & $10^{-14}$ & $10^{-14}$&$10^{-14}$&$10^{-14}$&$10^{-14}$&$10^{-14}$&$10^{-12}$ \\
  \hline 
  $ \qquad1$ & $10^{-12}$ & $10^{-13}$ & $10^{-13}$&$10^{-13}$&$10^{-13}$&$10^{-13}$&$10^{-13}$&$10^{-13}$&$10^{-13}$\\
  \hline 
  $ \qquad2$ & $10^{-10}$& $10^{-11}$ &$10^{-11}$&$10^{-11}$&$10^{-11}$&$10^{-11}$ &$10^{-11}$&$10^{-11}$&$10^{-11}$\\
    \hline 
  $ \qquad5$ & $10^{-8}$& $10^{-10}$ &$10^{-11}$&$10^{-12}$&$10^{-13}$&$10^{-10}$ &$10^{-10}$&$10^{-10}$&$10^{-10}$\\
    \hline 
  $ \qquad10$ & $10^{-6}$& $10^{-8}$ &$10^{-10}$&$10^{-11}$&$10^{-11}$&$10^{-9}$&$10^{-9}$&$10^{-9}$&$10^{-9}$ \\
  \hline  
 \end{tabular}
\caption{Comparison of the $O(N^{-6})$ expansion $\cF^{1/N}$ in \eqref{Ftilde} and \eqref{higherW} from topological recursion, and the finite $N$ $\cF^\text{finite}$ for $N=2,\dots,10$ from orthogonal polynomials. The $1/N^2$ expansion is most accurate for small $g^2_\text{YM}=\lambda/N$ and large $N$.}
\label{int1}
\end{center}
\end{table}

\section{Constraining the holographic correlator}
\label{relate}

We will now use the single constraint from the mass deformed free energy computed to all orders in $1/c$ and $1/\lambda$ in the previous section, as well as the single constraint from the relation to the known IIB S-matrix in the flat space limit as reviewed in Section \ref{strong0}, to fix all the coefficients shown in the strong coupling 't Hooft expansion \eqref{MellinIntro} of $\langle SSSS\rangle$. The result in Mellin space is
\es{MellinFinal}{
   \cM =&\frac1 c\left[\frac{8}{(s-2)(t-2)(u-2)}+\frac{120\zeta(3)}{\lambda^{\frac32}}   +\frac{630\zeta(5)}{\lambda^{\frac52}}\left[s^2+t^2+u^2-3\right]+O(\lambda^{-3})\right]\\
   &+\frac{1}{c^2}
   \left[\frac{5\sqrt{\lambda}}{8}+\mathcal{M}^{\text{SG}|\text{SG}}+\frac{15}{4} +O(\lambda^{-\frac32})\right]\\
   &+\frac{1}{c^3}
   \left[\frac{7\lambda^{\frac32}}{3072}\left[s^2+t^2+u^2 - 3  \right]+O(\lambda)\right]+O(c^{-4})\,,
}
where recall that $u=4-s-t$ and $\mathcal{M}^{\text{SG}|\text{SG}}$ has a more complicated $s,t,u$ dependence that we gave in \eqref{M1loop}.

\subsection{Genus-one from localization}
\label{1loop}

We start by using the integrated constraint \eqref{simpInt}. The LHS of this equation involves the integrals \eqref{I} for the position space expressions in \eqref{TIntro}, which are
\es{integrals}{
I[\cT^\text{SG}]=\frac{1}{32}\,,\qquad I[\cT^0]=-\frac{1}{40}\,,\qquad I[\cT^2]=-\frac{2}{35}\,,\qquad I[\cT^{\text{SG}|\text{SG}}]=\frac{5}{32}\,.
}
The first three expressions were computed in \cite{Binder:2019jwn}, while the last was computed in this work by evaluating the integral numerically to high precision using the explicit position space expression in \cite{Aprile:2017bgs}.\footnote{I thank Hynek Paul for sending me an explicit formula for this very complicated expression.} It is remarkable that such a complicated expression has such a simple integral, which hints at a simpler hidden structure.

The RHS of \eqref{simpInt} is derivatives of the mass deformed free energy evaluated at zero mass, which was computed to $O(N^{-6})$ in \eqref{largeLam}, and can be written in the $1/c$ expansion \eqref{cExp} using \eqref{cToN}. To the order we considered in \eqref{MellinIntro} we found
\es{c2}{
\frac{\cF}{c^2}=&\frac{1}{c}\left[\frac14-\frac{3\zeta(3)}{\lambda^{\frac32}}+\frac{45\zeta(5)}{4\lambda^{\frac52}}+O(\lambda^{-\frac72})\right]+\frac{1}{c^2}\left[-\frac{\sqrt{\lambda}}{64}+\frac{1}{16}+O(\lambda^{-\frac32})\right]\\
&+\frac{1}{c^3}\left[\frac{\lambda^{\frac32}}{24576}+O(\sqrt{\lambda})\right]+O(c^{-4})\,.
}
The integrated constraint \eqref{simpInt} then fixes the coefficients in \eqref{MellinIntro} as  
\es{fixLoc}{
R^4:&\qquad B^0_0=120\zeta(3)\,,\qquad \overline{B}_0^0=\frac58\,,\qquad \overline{\overline{B}}_0^0=0\,,\\
\text{SG}|\text{SG}:&\qquad \overline{B}^{\text{SG}|\text{SG}}_0=\frac{15}{4}\,,\\
D^4R^4:&\qquad \frac{45\zeta(5)}{4}=-\frac{B_0^2}{40}-\frac{2B_2^2}{35}\,,\qquad {\overline{B_0^2}}=-\frac{16{\overline{B_2^2}}}{7}\,,\qquad\frac{1}{24576}=-\frac{\overline{\overline{B_0^2}}}{40}-\frac{2\overline{\overline{B_2^2}}}{35}\,,\\
}
where the constraints on the genus-zero tree level coefficients $B_0^0$, $B_0^2$, and $B_2^2$ were already derived in this way in \cite{Binder:2019jwn}. The genus-one $R^4$ coefficient $ \overline{B}_0^0$ completes the CFT derivation of $R^4$ in the strong coupling 't Hooft limit, which receives no perturbative higher genus corrections. These higher genus $R^4$ corrections would scale as $\frac{\lambda^{2n-\frac72}}{c^n}$ for $n>2$, which do not appear in $\cF$ as verified to genus-four in \eqref{largeLam}. The coefficient $\overline{B}^{\text{SG}|\text{SG}}_0$ of the constant ambiguity at order $1/c^2$, along with the non-analytic 1-loop term $\cT^{\text{SG}|\text{SG}}$ in position space \cite{Alday:2017xua,Aprile:2017bgs} (or $\cM^{\text{SG}|\text{SG}}$ in Mellin space \cite{Alday:2018kkw}), completes the 1-loop supergravity term. To fix the $D^4R^4$ terms we must next consider the flat space limit.

\subsection{Genus-two from localization and string theory}
\label{2loop}

We can fix the leading large $s,t$ coefficient at each order by taking the flat space limit \eqref{Gotfst} of the Mellin amplitude \eqref{MellinIntro}, and comparing to the known IIB S-matrix in \eqref{A}. We find that
\es{fixFlat}{
R^4:&\qquad B^0_0=120\zeta(3)\,,\qquad \overline{B}_0^0=\frac58\,,\qquad \overline{\overline{B}}_0^0=0\,,\\
D^4R^4:&\qquad B_2^2={630\zeta(5)}\,,\qquad {\overline{B_2^2}}=0\,,\qquad \overline{\overline{B_2^2}}=\frac{7}{3072}\,,\\
}
where the constraints on the genus-zero coefficients $B_0^0$ and $B_2^2$ were already derived in this way in \cite{Goncalves:2014ffa}. The genus-one $R^4$ coefficient $\overline{B}_0^0$ agrees between \eqref{fixLoc} and \eqref{fixFlat}, which is a nontrivial check of AdS/CFT at genus-one (and a somewhat trivial check to all higher genus order which both methods say must vanish). For $D^4R^4$, we can combine \eqref{fixLoc} and \eqref{fixFlat} to fix 
\es{fixBoth}{
B_0^2=-1890\zeta(5)\,,\qquad \overline{B}_0^2=0\,,\qquad \overline{\overline{B}}_0^2=-\frac{7}{1024}\,,
}
which completes the derivation of the nonzero genus-two $D^4R^4$ term, which is in fact the leading order genus-two, i.e. $O(c^{-3})$, term in $\langle SSSS\rangle$. Since no other genus $D^4R^4$ terms appear in the IIB S-matrix, and no such terms, which would scale as $\frac{\lambda^{2n-\frac92}}{c^n}$ for $n\neq3$, appear in $\cF$ as verified to genus-four in \eqref{largeLam}, we have thus fixed the $D^4R^4$ term in the holographic correlator to all genus order. 

\subsection{Unprotected CFT data to order $O(c^{-3})$}
\label{data}

Now that $\langle SSSS\rangle$ has been fixed to the order shown in \eqref{MellinFinal}, we can use it to extract any CFT data to this order that we like. For instance, we find the anomalous dimensions $\gamma_j$ of the unique lowest twist even spin $j$ double-trace operators $[S\partial_{\mu_1}\dots\partial_{\mu_j}S]$ to be
\es{anom}{
\gamma_j=&\frac{1}{c}\left[-\frac{24}{(j+1)(j+6)}-\frac{4320\zeta(3)}{7\lambda^{\frac32}}\delta_{j,0}-\frac{\zeta(5)}{\lambda^{\frac52}}\left[{30600}\delta_{j,0}+\frac{201600}{11}\delta_{j,2}\right]+O(\lambda^{-3})\right]\\
&+\frac{1}{c^2}\left[-\frac{45\sqrt{\lambda}}{14}\delta_{j,0}+\frac{24 \left(7 j^4+74 j^3-553 j^2-4904 j-3444\right)}{(j-1) (j+1)^3 (j+6)^3
   (j+8)}-\frac{135}{7}\delta_{j,0}\right]\\
   &+\frac{1}{c^3}\left[-{\lambda^{\frac32}}\left[\frac{85}{768}\delta_{j,0}+\frac{35}{528}\delta_{j,2}\right]+O(\lambda)\right]+O(c^{-4})\,,
}
where the three $O(c^{-1})$ terms were computed in \cite{Arutyunov:2000ku,DHoker:1999mic}, \cite{Goncalves:2014ffa}, and \cite{Binder:2019jwn}, respectively. Contact terms with $n$-derivatives only contribute to operators up to spin $n/2-4$, as explained in \cite{Heemskerk:2009pn}. The 1-loop supergravity term at order $1/c^2$ was originally computed for all spins in \cite{Aprile:2017bgs}, where it was conjectured that $\overline{B}_0^{\text{SG}|\text{SG}}$ was zero so that the $1/c^2$ term would be analytic in spin down to $j=0$. In fact, analyticity in spin is only expected for $j>0$ in strongly coupled $\mathcal{N}=4$ SYM at 1-loop order \cite{Caron-Huot:2018kta},\footnote{At finite $N$, the theory is analytic for all $j\geq0$, but in the strong coupling expansion the analyticity worsens at each order in $1/c$. For instance, while the theory is still analytic for all $j\geq0$ at tree level, at 1-loop it is now only analytic for $j>0$ \cite{Caron-Huot:2018kta}.} and the fact that $j=0$ differs from $j>0$ is a striking validation of this fact. For $j>0$, the contributions to the anomalous dimension from the loop amplitudes $\cM^{\text{SG}|R^4_\text{genus-0}}$ and  $\cM^{\text{SG}|D^4R^4_\text{genus-0}}$ have also been computed in \cite{Alday:2018pdi,Binder:2019jwn}, but the ambiguities needed to compute all spins have not yet been fixed. For higher twist there are many degenerate double-trace operators, so one would need to compute many different half-BPS correlators to determine their anomalous dimensions \cite{Alday:2017xua,Aprile:2017bgs}.

\section{Conclusion}
\label{conc}

In this work we computed the four point function $\langle SSSS\rangle$ of the superprimary of the stress tensor multiplet in $\mathcal{N}=4$ SYM with gauge group $SU(N)$ in the strong coupling 't Hooft limit at large $c\sim N^2$ and large $\lambda$, which is holographically dual to scattering of supergravitons in IIB string theory on $AdS_5\times S^5$ expanded for small $g_s$ and $\ell_s$. The integral of $\langle SSSS\rangle$ was related in \cite{Binder:2019jwn} to derivatives $\partial_m^2\partial_\tau\partial_{\bar\tau}F$ of the mass deformed $\mathcal{N}=2^*$ free energy $F$ on $S^4$, which can be expressed using supersymmetric localization as an expectation value in a free Gaussian matrix model \cite{Pestun:2007rz}. This quantity was previously only known to leading order in the 't Hooft limit \cite{Russo:2013kea}. Our main technical result was a derivation of $\partial_m^2\partial_\tau\partial_{\bar\tau}F$ to any order in $1/N$ at finite or perturbative $\lambda$ using topological recursion \cite{Eynard:2004mh,Eynard:2008we}, which we verified at finite $N$ and $\lambda$ using orthogonal polynomials \cite{mehta1981}. 

We used this any order result to fix the $R^4$ correction to $\langle SSSS\rangle$ to all orders in the genus expansion, of which only the genus-zero and one are nonzero. We were also able to fix the constant ambiguity in the 1-loop supergravity contribution, which completes the derivation of this term as initiated in \cite{Alday:2017xua,Aprile:2017bgs,Alday:2018kkw}, and shows that analyticity in spin for the lowest twist anomalous dimension fails for zero spin, as expected from the Lorentzian inversion formula \cite{Caron-Huot:2017vep}. We also used the known IIB S-matrix, which is related to the flat space limit of $\langle SSSS\rangle$, to constrain this correlator. This gave the same result for $R^4$, which is the first genus-one check of AdS$_5/$ CFT$_4$ for local operators that could not be determined from genus-zero. By combining localization and the flat space limit we then fixed $D^4R^4$ to all orders in the genus expansion, and verified that only the genus-zero and two contributions are nonzero. This genus-two term scales as $\lambda^{\frac32}c^{-3}$, and is the first correction to $\langle SSSS\rangle$ computed at $O(c^{-3})$.

There are many future directions to this work. One could generalize the any order in $1/N$ localization computation to SYM with gauge group $SO(N)$ or $Sp(N)$, whose matrix model is also known \cite{Pestun:2007rz}. While these other gauge group lead to the same holographic correlator for tree level supergravity, they likely differ for the higher order corrections considered in this work. One could also generalize the computation to correlators $\langle S_2S_2 S_pS_p\rangle$ of two stress tensor multiplets and two single-trace half-BPS multiplets whose superprimary has dimension $p>2$, which was related in \cite{Binder:2019jwn} to derivatives $\partial_m^2\partial_{\tau_p}\partial_{\bar\tau_p}F$ of the sphere free energy deformed by the coupling $\tau_p$ to the top components of these other half-BPS multiplets. The  partition function (not counting instantons) for this deformation was computed using localization in terms of a matrix model in \cite{Gerchkovitz:2016gxx}, and resembles the $p=2$ matrix model considered in this work except that the Gaussian term in \eqref{N2starMatrixModel} is replaced by
\es{pq}{
e^{-\frac{8\pi^2 N\sum_i a_i^2}{\lambda}}\to e^{-\frac{8\pi^2 N\sum_i a_i^2}{\lambda}}e^{i\frac{\pi^{p/2} (\tau'_p-\bar\tau'_p)\sum_i a_i^p}{\lambda}}\,,
}
where the coefficient $\tau_p'$ in general differs from the coupling $\tau_p$ due to operator mixing on $S^4$. While topological recursion in fact applies to any polynomial potential \cite{Eynard:2004mh,Eynard:2008we} such as this one, it is difficult to resolve this operator mixing beyond the leading order in $1/N^2$ unmixing that was done in \cite{Rodriguez-Gomez:2016ijh,Binder:2019jwn}. If this quantity could be computed, then it could be used to fix the constant ambiguity in the 1-loop supergravity contribution to $\langle S_2S_2 S_pS_p\rangle$, which was explicitly derived for $p=3$ in \cite{Aprile:2017qoy}. The other linear in $s,t$ ambiguity that appears in this term could then be fixed from the flat space limit, since it is the same order in large $s,t$ as the non-analytic terms.

One could also consider new integrated constraints that come from four mass derivatives of $F$, or derivatives in terms of the squashing parameter $b$ for the free energy $F_b$ on the squashed sphere, which was also computed in terms of a matrix model using localization in \cite{Hama:2012bg}. All these quantities take the form of expectation values in the undeformed free Gaussian matrix model for $\mathcal{N}=4$ SYM, and so could be computed to any order in $1/N$ using the methods of this paper. These additional constraints could then be used to fix $\langle SSSS\rangle$ up to genus-three.\footnote{We do not expect to be able to derive more than 3 constraints that are independent in the large $N$ limit, because otherwise we would be able to use localization to fix the unprotected $D^8R^4$ term.} They could also be used to fix the ambiguities in the 1-loop term $\cM^{\text{SG}|R^4_\text{genus-0}}$ with one supergravity vertex and one genus-zero $R^4$ vertex \cite{Alday:2018pdi,Alday:2018kkw}, which is asymptotically degree 4 in $s,t,u$ and so contains four polynomial ambiguities that must be fixed.

In this work we considered the strong coupling 't Hooft limit, but one could also consider the holographic limit where $N\to\infty$ and $\tau=\frac{\theta}{2\pi}+\frac{4\pi i}{g_\text{YM}^2}$ is finite, which corresponds to the small $\ell_s$ and finite $\tau_s=\chi_s+ig_s^{-1}$ expansion in the IIB S-matrix, where $\chi_s$ is the axion coupling. The coefficients for each expansion must be $SL(2,Z)$ invariants of $\tau$ and $\tau_s$, respectively, and indeed the coefficients of the protected $R^4$, $D^4R^4$, and $D^6R^4$ in the IIB S-matrix involve non-holomorphic Eisenstein series \cite{Green:1997as,Green:1998by,Green:1999pu,Green:2005ba}. In \cite{Binder:2019jwn}, the flat space limit was used to show that the coefficient of the $R^4$ term in $\langle SSSS\rangle$ must also be an Eisenstein series. To derive this from the mass deformed partition function one would need to consider the contribution of the instantons in the Nekrasov partition function to the matrix model expectation value. It would nice to see if this is possible to compute with our methods to any order in $1/N$.

Lastly, while the application of integrated constraints and localization to holographic correlators has been perturbative in this paper and the original work \cite{Binder:2019jwn}, these relations are in fact non-perturbative, and so could be applied to the numerical bootstrap for $\mathcal{N}=4$ SYM \cite{Beem:2013qxa,Beem:2016wfs}. For this purpose, the finite $N$ formula for the perturbative part of the mass deformed free energy, as derived using orthogonal polynomials in this work, will be especially useful, especially if one could augment it with a similar formula for the contribution from the Nekrasov partition function. These constraints could allow one to impose the values of $\tau$ and $\bar\tau$ in the numerical bootstrap for finite $N$, just as $N$ was imposed in the original studies \cite{Beem:2013qxa,Beem:2016wfs} using the conformal anomaly $c$, and thereby solve $\mathcal{N}=4$ SYM numerically for all $\tau$, $\bar\tau$ and $N$.

\section*{Acknowledgments} 

I thank Ofer Aharony, Michael Green, Masazumi Honda, Hynek Paul, Eric Perlmutter, Silviu Pufu, Rohit Reghupathy, Adar Sharon, Yifan Wang, Congkau Wen, and Xinan Zhou for useful discussions, and Damon Binder, Eric Perlmutter, Silviu Pufu, Yifan Wang, and Xi Yin for previous collaborations on related subjects.  I am supported by the Zuckerman STEM Leadership Fellowship. I also thank the organizers of ``Bootstrap 2019'' and Perimeter Institute for Theoretical Physics, as well as the Aspen Institute for Physics, which is supported by National Science Foundation grant PHY-1607611, for hospitality during the course of this work.

\appendix

\section{Comparing $\cM^{\text{SG}|\text{SG}}$ and $\cT^{\text{SG}|\text{SG}}$}
\label{1loopAp}
In this appendix we check that $\cM^{\text{SG}|\text{SG}}$ as defined in \eqref{M1loop} with $d_{mn}$ and $C$ fixed in \eqref{dc} is equivalent to $\cT^{\text{SG}|\text{SG}}$ as given in \cite{Aprile:2017bgs}. In Appendix D of \cite{Alday:2018kkw}, it was checked that the $U^2\log U$ for finite $V$ term in $\cM^{\text{SG}|\text{SG}}$ yields the same lowest twist double-trace anomalous dimension for spin $j>0$ as given in \cite{Aprile:2017bgs}. Since this check was only done for $j>0$, it was not sensitive to the constant terms $d_{mn}$ and $C$ that only contributes $j=0$ data. Indeed, in \cite{Alday:2018kkw} $d_{mn}$ and $C$ were not specified, so it was impossible to compare the results in  \cite{Alday:2018kkw} and \cite{Aprile:2017bgs} for $j=0$.

Our strategy here is to extract the leading $U^2\log U\log V$ term, which contributes to $j=0$ CFT data, from $\cM^{\text{SG}|\text{SG}}$ and compare it directly to $\cT^{\text{SG}|\text{SG}}$. From the Mellin transform \eqref{mellinDefD}, we see that to extract this term we must take the $s=4$ and $t=4$ poles. After taking these poles and performing the sums over $m$ and $n$, which are finite due to the $d_{mn}$ term, we find that the $U^2\log U\log V$ term is
\es{check}{
\cT^{\text{SG}|\text{SG}}\vert_{U^2\log U\log V}= -171 + 8 \pi^2\,.
}
This can be matched with the relevant term in the small $U,V$ expansion of $\cT^{\text{SG}|\text{SG}}$ that can be extracted from the explicit expression in \cite{Aprile:2017bgs}.

\section{Resolvents}
\label{res}
In this appendix we give the explicit results for the resolvents defined in \eqref{W} that we need to compute $\cF$ to genus 4, which we computed using the recursion method described in the main text. We already gave $W_0^1$ and $W_0^2$ in \eqref{W01} and \eqref{W20}, respectively. The higher ``genus'' one-body resolvents are
\es{moreRes}{
W_1^1=&\left[\frac{4\pi^2y_1^2}{\lambda}-1\right]^{-\frac52}\frac{\pi }{8 \sqrt{\lambda }}\,,\\
W_2^1=&\left[\frac{4\pi^2y_1^2}{\lambda}-1\right]^{-\frac{11}{2}}\left[\frac{21 \pi }{512 \sqrt{\lambda }}+\frac{21 \pi ^3 y_1^2}{32 \lambda ^{3/2}}\right]\,,\\
W_3^1=&\left[\frac{4\pi^2y_1^2}{\lambda}-1\right]^{-\frac{17}{2}}\left[\frac{869 \pi }{16384 \sqrt{\lambda }}+\frac{1485 \pi ^5 y_1^4}{128 \lambda ^{5/2}}+\frac{3069 \pi ^3 y_1^2}{1024 \lambda ^{3/2}}\right]\,,\\
W_4^1=&\left[\frac{4\pi^2y_1^2}{\lambda}-1\right]^{-\frac{23}{2}}\left[\frac{334477 \pi }{2097152 \sqrt{\lambda }}+\frac{225225 \pi ^7 y_1^6}{512 \lambda ^{7/2}}+\frac{1957527 \pi ^5 y_1^4}{8192 \lambda ^{5/2}}+\frac{1314027 \pi ^3 y_1^2}{65536 \lambda ^{3/2}}\right]\,,\\
}
and their inverse Laplace transforms at $2i\omega$ are
\es{IL1}{
\cL^{-1}[W_1^1(y_1)](2i\omega)=&\frac{\lambda  \omega ^2 J_2(\frac{\sqrt{\lambda } \omega }{\pi
   })}{48 \pi ^2}\,,\\
   \cL^{-1}[W_2^1(y_1)](2i\omega)=&\frac{\lambda ^2 \omega ^4 J_4(\frac{\sqrt{\lambda } \omega }{\pi
   })}{1280 \pi ^4}-\frac{\lambda ^{5/2} \omega ^5
   J_5(\frac{\sqrt{\lambda } \omega }{\pi })}{9216 \pi ^5}\,,\\
   \cL^{-1}[W_3^1(y_1)](2i\omega)=&-\frac{\lambda ^{7/2} \omega ^7 J_7(\frac{\sqrt{\lambda } \omega }{\pi
   })}{122880 \pi ^7}+\frac{\lambda ^{5/2} \omega ^5
   J_7(\frac{\sqrt{\lambda } \omega }{\pi })}{2048 \pi
   ^5}+\frac{\lambda ^4 \omega ^8 J_8(\frac{\sqrt{\lambda } \omega
   }{\pi })}{2654208 \pi ^8}-\frac{\lambda ^3 \omega ^6
   J_8(\frac{\sqrt{\lambda } \omega }{\pi })}{28672 \pi ^6}\,,\\
   \cL^{-1}[W_4^1(y_1)](2i\omega)=&\frac{\lambda ^3 \omega ^6}{{178362777600 \pi ^{11}}} \textstyle\left(1080 \left(7 \pi  \lambda ^2 \omega
   ^4-1984 \pi ^3 \lambda  \omega ^2+100800 \pi ^5\right)
   J_{10}(\frac{\sqrt{\lambda } \omega }{\pi })\right.\\
   &\left.+\sqrt{\lambda }
   \omega  \left(-175 \lambda ^2 \omega ^4+92016 \pi ^2 \lambda  \omega
   ^2-5443200 \pi ^4\right) J_{11}(\frac{\sqrt{\lambda } \omega }{\pi
   })\right)\,.
}
The higher ``genus'' two-body resolvents are
\es{moreRes2}{
W_1^2=&\left[\frac{4\pi^2y_1^2}{\lambda}-1\right]^{-\frac72}\left[\frac{4\pi^2y_2^2}{\lambda}-1\right]^{-\frac72}  \frac{1}{16 \lambda ^4} \Big[5 \pi ^2 \lambda ^3-4 \pi ^4 \lambda ^2 \left(y_1^2-13 y_2 y_1+y_2^2\right)\\
&+16 \pi ^6 \lambda 
   \left(y_1^4-13 y_2 y_1^3-13 y_2^2 y_1^2-13 y_2^3 y_1+y_2^4\right)+64 \pi ^8 y_1 y_2 \left(5 y_1^4+4
   y_2 y_1^3+3 y_2^2 y_1^2+4 y_2^3 y_1+5 y_2^4\right) \Big]\,,\\
   }
   \es{moreRes3}{
W_2^2=&\left[\frac{4\pi^2y_1^2}{\lambda}-1\right]^{-\frac{13}{2}}\left[\frac{4\pi^2y_2^2}{\lambda}-1\right]^{-\frac{13}{2}}\frac{\pi^2}{256\lambda^8}\Big[-106 \lambda ^7-\pi ^2 \lambda ^6 \left(547 y_1^2+2882 y_2 y_1+547 y_2^2\right)\\
&-4 \pi ^4 \lambda ^5
   \left(302 y_1^4-5531 y_2 y_1^3-9842 y_2^2 y_1^2-5531 y_2^3 y_1+302 y_2^4\right)\\
   &+16 \pi ^6 \lambda ^4
   \left(203 y_1^6-7624 y_2 y_1^5-7673 y_2^2 y_1^4+3168 y_2^3 y_1^3-7673 y_2^4 y_1^2-7624 y_2^5 y_1+203
   y_2^6\right)\\
   &-64 \pi ^8 \lambda ^3 \left(106 y_1^8-5773 y_2 y_1^7-6320 y_2^2 y_1^6+8103 y_2^3
   y_1^5+8328 y_2^4 y_1^4+8103 y_2^5 y_1^3-6320 y_2^6 y_1^2\right.\\
&\left.   -5773 y_2^7 y_1+106 y_2^8\right)+256 \pi
   ^{10} \lambda ^2 \left(21 y_1^{10}-2352 y_2 y_1^9-2629 y_2^2 y_1^8+8434 y_2^3 y_1^7+7745 y_2^4
   y_1^6+6906 y_2^5 y_1^5\right.\\
   &\left.+7745 y_2^6 y_1^4+8434 y_2^7 y_1^3-2629 y_2^8 y_1^2-2352 y_2^9 y_1+21
   y_2^{10}\right)+1024 \pi ^{12} \lambda  y_1 y_2 \left(399 y_1^{10}+462 y_2 y_1^9\right.\\
   &\left.-4011 y_2^2
   y_1^8-3664 y_2^3 y_1^7-3317 y_2^4 y_1^6-3342 y_2^5 y_1^5-3317 y_2^6 y_1^4-3664 y_2^7 y_1^3-4011 y_2^8
   y_1^2+462 y_2^9 y_1\right.\\
   &\left.+399 y_2^{10}\right)+49152 \pi ^{14} y_1^3 y_2^3 \left(63 y_1^8+56 y_2 y_1^7+49
   y_2^2 y_1^6+52 y_2^3 y_1^5+55 y_2^4 y_1^4+52 y_2^5 y_1^3+49 y_2^6 y_1^2\right.\\
   &\left.+56 y_2^7 y_1+63 y_2^8\right)\Big]\,,\\
   }
      \es{moreRes32}{
W_3^2=&\left[\frac{4\pi^2y_1^2}{\lambda}-1\right]^{-\frac{19}{2}}\left[\frac{4\pi^2y_2^2}{\lambda}-1\right]^{-\frac{19}{2}}\frac{\pi^2}{4096\lambda^{12}}\Big[5165 \lambda ^{11}+4 \pi ^2 \left(26264 y_1^2+69583 y_2 y_1+26264 y_2^2\right) \lambda
   ^{10}\\
   &+2 \pi ^4 \left(182681 y_1^4-1229348 y_2 y_1^3-3301058 y_2^2 y_1^2-1229348 y_2^3
   y_1+182681 y_2^4\right) \lambda ^9-8 \pi ^6 \left(43363 y_1^6\right.\\
   &\left.-3247129 y_2
   y_1^5-2154755 y_2^2 y_1^4+5886474 y_2^3 y_1^3-2154755 y_2^4 y_1^2-3247129 y_2^5
   y_1+43363 y_2^6\right) \lambda ^8\\
   &+32 \pi ^8 \left(66530 y_1^8-4848035 y_2
   y_1^7-5522929 y_2^2 y_1^6+13273587 y_2^3 y_1^5+18753102 y_2^4 y_1^4+13273587 y_2^5
   y_1^3\right.\\
   &\left.-5522929 y_2^6 y_1^2-4848035 y_2^7 y_1+66530 y_2^8\right) \lambda ^7-128 \pi
   ^{10} \left(50475 y_1^{10}-4865970 y_2 y_1^9-5594975 y_2^2 y_1^8\right.\\
   &\left.+22854785 y_2^3
   y_1^7+22473748 y_2^4 y_1^6+9367026 y_2^5 y_1^5+22473748 y_2^6 y_1^4+22854785 y_2^7
   y_1^3-5594975 y_2^8 y_1^2\right.\\
   &\left.-4865970 y_2^9 y_1+50475 y_2^{10}\right) \lambda ^6+512 \pi
   ^{12} \left(24774 y_1^{12}-3255747 y_2 y_1^{11}-3856071 y_2^2 y_1^{10}+24719535 y_2^3
   y_1^9\right.\\
   &\left.+24707885 y_2^4 y_1^8+5373820 y_2^5 y_1^7+5186920 y_2^6 y_1^6+5373820 y_2^7
   y_1^5+24707885 y_2^8 y_1^4+24719535 y_2^9 y_1^3\right.\\
   &\left.-3856071 y_2^{10} y_1^2-3255747
   y_2^{11} y_1+24774 y_2^{12}\right) \lambda ^5-2048 \pi ^{14} \left(7023
   y_1^{14}-1401654 y_2 y_1^{13}-1707549 y_2^2 y_1^{12}\right.\\
   &\left.+17443791 y_2^3 y_1^{11}+17426448
   y_2^4 y_1^{10}-2018205 y_2^5 y_1^9-1319170 y_2^6 y_1^8-404360 y_2^7 y_1^7-1319170
   y_2^8 y_1^6\right.\\
   &\left.-2018205 y_2^9 y_1^5+17426448 y_2^{10} y_1^4+17443791 y_2^{11}
   y_1^3-1707549 y_2^{12} y_1^2-1401654 y_2^{13} y_1+7023 y_2^{14}\right) \lambda ^4\\
   &+8192
   \pi ^{16} \left(869 y_1^{16}-352187 y_2 y_1^{15}-441465 y_2^2 y_1^{14}+7809017 y_2^3
   y_1^{13}+7802545 y_2^4 y_1^{12}-5171532 y_2^5 y_1^{11}\right.\\
   &\left.-4601146 y_2^6 y_1^{10}-3974690
   y_2^7 y_1^9-4031390 y_2^8 y_1^8-3974690 y_2^9 y_1^7-4601146 y_2^{10} y_1^6-5171532
   y_2^{11} y_1^5\right.\\
   &\left.+7802545 y_2^{12} y_1^4+7809017 y_2^{13} y_1^3-441465 y_2^{14}
   y_1^2-352187 y_2^{15} y_1+869 y_2^{16}\right) \lambda ^3\\
   &+32768 \pi ^{18} y_1 y_2
   \left(39325 y_1^{16}+50732 y_2 y_1^{15}-2022801 y_2^2 y_1^{14}-2020136 y_2^3
   y_1^{13}+3542369 y_2^4 y_1^{12}\right.\\
   &\left.+3237432 y_2^5 y_1^{11}+2923150 y_2^6 y_1^{10}+2984420
   y_2^7 y_1^9+3023010 y_2^8 y_1^8+2984420 y_2^9 y_1^7+2923150 y_2^{10} y_1^6\right.\\
   &\left.+3237432
   y_2^{11} y_1^5+3542369 y_2^{12} y_1^4-2020136 y_2^{13} y_1^3-2022801 y_2^{14}
   y_1^2+50732 y_2^{15} y_1+39325 y_2^{16}\right) \lambda ^2\\
   &+4718592 \pi ^{20} y_1^3
   y_2^3 \left(6435 y_1^{14}+6424 y_2 y_1^{13}-32197 y_2^2 y_1^{12}-29686 y_2^3
   y_1^{11}-27175 y_2^4 y_1^{10}-27660 y_2^5 y_1^9\right.\\
   &\left.-28040 y_2^6 y_1^8-28040 y_2^7
   y_1^7-28040 y_2^8 y_1^6-27660 y_2^9 y_1^5-27175 y_2^{10} y_1^4-29686 y_2^{11}
   y_1^3-32197 y_2^{12} y_1^2\right.\\
   &\left.+6424 y_2^{13} y_1+6435 y_2^{14}\right) \lambda +188743680
   \pi ^{22} y_1^5 y_2^5 \left(429 y_1^{12}+396 y_2 y_1^{11}+363 y_2^2 y_1^{10}+372 y_2^3
   y_1^9+381 y_2^4 y_1^8\right.\\
   &\left.+376 y_2^5 y_1^7+371 y_2^6 y_1^6+376 y_2^7 y_1^5+381 y_2^8
   y_1^4+372 y_2^9 y_1^3+363 y_2^{10} y_1^2+396 y_2^{11} y_1+429 y_2^{12}\right)\Big]\,.\\
}
Note that all these expressions factorize in terms of $y_1,y_2$. Their inverse Laplace transforms at $2i\omega,-2i\omega$ can then be easily computed to get
\es{IL2}{
\cL^{-1}[W_1^2&(y_1,y_2)](2i\omega,-2i\omega)=-\frac{-4 \pi  \lambda ^{3/2} \omega ^4 J_1(\frac{\sqrt{\lambda }
   \omega }{\pi }) J_2(\frac{\sqrt{\lambda } \omega }{\pi
   })+\lambda ^2 \omega ^5 J_1(\frac{\sqrt{\lambda } \omega
   }{\pi })^2+\lambda ^2 \omega ^5 J_2(\frac{\sqrt{\lambda }
   \omega }{\pi })^2}{192 \pi ^4 \omega }\,,\\
   \cL^{-1}[W_2^2&(y_1,y_2)](2i\omega,-2i\omega)=-\frac{\lambda ^{3/2} \omega ^3}{92160
   \pi ^6} \Big[\textstyle2 \sqrt{\lambda } \omega 
   \left(\lambda  \omega ^2+144 \pi ^2\right) J_4(\frac{\sqrt{\lambda
   } \omega }{\pi })^2\\
   &\textstyle+96 \pi  \left(2 \pi ^2-\lambda  \omega
   ^2\right) J_5(\frac{\sqrt{\lambda } \omega }{\pi })
   J_4(\frac{\sqrt{\lambda } \omega }{\pi })+\sqrt{\lambda }
   \omega  \left(7 \lambda  \omega ^2-24 \pi ^2\right)
   J_5(\frac{\sqrt{\lambda } \omega }{\pi })^2\Big]\,,\\
   \cL^{-1}[W_3^2&(y_1,y_2)](2i\omega,-2i\omega)=-\frac{\lambda  \omega ^2}{92897280
   \pi ^9} \Big[\textstyle12 \pi  \lambda  \omega ^2 \left(57 \lambda
   ^2 \omega ^4-7640 \pi ^2 \lambda  \omega ^2+222720 \pi ^4\right)
   J_8(\frac{\sqrt{\lambda } \omega }{\pi })^2\\
   &\textstyle+24 \left(-31
   \pi  \lambda ^3 \omega ^6+12510 \pi ^3 \lambda ^2 \omega ^4-1001520 \pi
   ^5 \lambda  \omega ^2+21772800 \pi ^7\right)
   J_7(\frac{\sqrt{\lambda } \omega }{\pi })^2\\
   &\textstyle+5 \sqrt{\lambda
   } \omega  \left(7 \lambda ^3 \omega ^6-5952 \pi ^2 \lambda ^2 \omega
   ^4+600192 \pi ^4 \lambda  \omega ^2-14948352 \pi ^6\right)
   J_8(\frac{\sqrt{\lambda } \omega }{\pi })
   J_7(\frac{\sqrt{\lambda } \omega }{\pi })\Big]\,.\\
}

\bibliographystyle{ssg}
\bibliography{N41loop}

\end{document}